\documentstyle{article}  
\newcommand{\be}{\begin{equation}}
\newcommand{\eq}{\end{equation}}
\newcommand{\ov}{\overline}

\newcommand{\la}{\langle}
\newcommand{\ra}{\rangle}

\newcommand{\Zzahl}{ {\bf Z} }

\newcommand{\Rzahl}{ {\bf R} }

\newcommand{\wt}{\widetilde}
\newcommand{\wh}{\widehat}
\newcommand{\fs}[1]{\mbox{\scriptsize \bf #1}}

\begin{document}

\pagestyle{empty}
\renewcommand{\thefootnote}{\fnsymbol{footnote}}

\vspace{-3cm} {\normalsize \hfill MS-TPI-93-09}      \\[25mm]

\begin{center}
{\LARGE
Orbifold Compactifications with Continuous Wilson Lines
}
\end{center}

\vspace{1cm}
\begin{center}
{\large
Thomas Mohaupt\footnote{New address from 1 november 1993:
DESY--IfH Zeuthen, Platanenallee 6, D--15738 Zeuthen.
(Postal address: P.O. Box, D--15735 Zeuthen. E-mail: mohaupt at
ifh.de.)
} \\[0.4cm]
        Institut f\"ur Theoretische Physik I,
        Universit\"at M\"unster\\[0.4cm]
        Wilhelm-Klemm-Str.~9, D-48149 M\"unster, Germany \\[0.4cm]
October 27, 1993\footnote{Corrected: November 8, 1993.}
}
\end{center}

\vspace{1cm}
\begin{center}
{\bf Abstract}: \\[0.4cm]
\end{center}

We identify the untwisted moduli of heterotic
orbifold compactifications for the case, when the gauge twist
is realized by a rotation. The Wilson lines are found to
have both continuous and discrete parts. For the case of
the standard $\Zzahl_{3}$ orbifold we classify all possibilities
of breaking the gauge group $E(6) \otimes SU(3)$ by
nine of the eighteen Wilson moduli and by additional discrete Wilson
lines.

\setcounter{page}{0}
\newpage

\renewcommand{\thefootnote}{\arabic{footnote}}
\setcounter{footnote}{0}

\pagestyle{myheadings}
\markright{T. Mohaupt,
           Orbifold Compactifications with Continuous Wilson Lines}

\section{Introduction}

The compactification of the heterotic string theory \cite{GSW}
from ten to four dimensions can be realized in a lot of different
ways \cite{Kaku}. A generic feature of all these schemes is the
appearence of contiuous deformation parameters, called moduli
\cite{DDV,FerThe}.
The moduli spaces of toroidal \cite{Nar,NSW}
and orbifold compactifications \cite{DHVW1,DHVW2,NSV}
have been studied extensively during
the last years. This is due to the fact that these models are
exactly solvable and lead, at least for some orbifolds, to
semirealistic physics \cite{INQ,IKNQ}.
Besides being of some phenomenological
interest, they are a good laboratory for ideas about string
theory that may later be generalized to more general
compactifications or internal conformal field theories.

Perhaps the most prominent point of interest is
target space modular
invariance, a symmetry of the moduli space,
which is a genuine stringy phenomenon \cite{GRV,LMN,FerThe}.
Since this modular group contains
inversions of the radius of the internal manifold, it indicates
the existence of a fundamental length scale. On the other hand,
demanding the absence of modular anomalies puts severe
restrictions on the effective Lagrangean describing the
low (compared to Planck scale) energy regime \cite{FLST,IbaLue}.

Another point of interest is the dependence of the gauge group
on the moduli \cite{Gin,Moh}. This is not only useful
for identifying models
with standard model or GUT gauge groups, but also linked with
target space modular invariance, because
points of maximally extended symmetry are often fixed points of
the modular group \cite{GinRev}. It has been argued that both
the continuous, finite dimensional gauge group and the
discrete modular group are remnants of the full, (spontanously broken)
infinite dimensional, continuous string symmetry group \cite{Giv}.

It is therefore surprising that a big part of the
orbifold moduli space has not yet
been explored systematically: The moduli associated with continuous
Wilson lines. These moduli
are known to exist, when the gauge twist is not realized by a shift,
but by a rotation \cite{IMNQ}.
In this paper we will show that it is possible
in the case of $\Zzahl_{N}$ orbifolds to determine all
(untwisted\footnote{The so called twisted moduli, which are associated
with the blowing up of orbifold singularities will not be studied
here.})
moduli including the Wilson moduli. Our approach will be similar
to that used in \cite{IMNQ}, namely we will analyze the constraint
that the twist must be an automorphism of the momentum
lattice of the underlying toroidal model. In \cite{IMNQ}
this was applied to symmetric orbifolds with gauge twist
realized by a shift and to general asymmetric orbifolds,
whereas the case which we will study (called symmetric
orbifolds with non abelian embedding in \cite{IMNQ}\footnote{Such
orbifolds are called asymmetric by other authors, because they
generically have $(0,2)$ world sheet supersymmetry only. })
was only mentioned
but not treated explicitly. Note also that in \cite{IMNQ} no
formula was derived for the number of moduli. Since this has been
shown to be possible more recently in \cite{EJL} for bosonic orbifolds,
we will try to do the same in the heterotic case.
Our second task will be to characterize the moduli dependence
of the gauge group. This will be done using the results for
toroidal models derived in \cite{Moh}.

The organisation of this paper is as follows: In the next section
we will recall some background material. Then the
constraints resulting from demanding consistency between a given
twist of the momentum lattice and the background fields will
be worked out. By analyzing the constraints we will learn,
which parts of the Wilson lines can be varied continuously
and which are discrete. As a byproduct we will show that
discrete (``quantized'') background fields reflect the remaining freedom
of selecting different twists of the momentum lattice, when
a target space twist\footnote{The terminology will be explained in
the next section.} has already been chosen.
Then we will define the gauge twist in terms of the target space twist
and derive formulas for the number of moduli. The general
results will then be illustrated taking the standard $\Zzahl_{3}$
Orbifold as an example. We will classify
all possibilities of breaking the gauge group $E(6) \otimes SU(3)$
by nine of the eighteen Wilson moduli and by additional discrete
Wilson lines.

\section{Background material}

Toroidal orbifold models are constructed by modding out
a discrete symmetry of a toroidal compactification
of the ten dimensional heterotic string theory.
We must therefore recall some elements of the Narain model
\cite{Nar,NSW}.

\subsection{Toroidal compactifications}

The Narain model desribes strings moving in a space--time
${\bf M}^{10 - d} \times {\bf T}^{d}$, where
${\bf M}^{10 - d}$ is a $(10 - d)$--dimensional Minkowski space and
${\bf T}^{d}$  is a $d$--dimensional torus, defined by
\be     {\bf T}^{d} = {\bf R}^{d} /  \Lambda,
\eq
where $\Lambda$ is a $d$--dimensional lattice.
In the following we will need a lattice basis
${\bf e}_{i}$ of $\Lambda$,
\be     \Lambda = \la {\bf e}_{i} | i = 1, \ldots, d \ra_{\fs{Z}}
        := \{ m^{i}{\bf e}_{i} | m^{i} \in {\bf Z} \}.
\eq
Inequivalent choices of $\Lambda$ can be parametrized by the
lattice metric
\be     G_{ij} := {\bf e}_{i} \cdot {\bf e}_{j} .
\eq
They correspond to different background metrics
in the $d$ internal directions.
But there are further continuous deformations:
One can also introduce a constant antisymmetric
background field $B_{ij}$, sometimes called axion
and a constant background gauge field
of the ten--dimensional gauge group, which can be paramatrized by
six vectors ${\bf A}_{i} \in {\bf R}^{16}$, called Wilson lines
\cite{NSW}.
A basis of the Hilbert space of the model is provided by
acting with transverse oszillators on the
ground states
\be     |({\bf K}, {\bf k}_{L}; {\bf k}_{R}) \ra
\eq
of the different charge/winding/momentum
sectors\footnote{From the four--dimensional point of view
the components of all thesese vectors are charge--like
quantum numbers.}.
The vectors $({\bf K}, {\bf k}_{L}; {\bf k}_{R})$ are
constrained to form an even lattice $\Gamma_{16+d;d}$, which is
selfdual with respect
to an indefinite bilinear form with signature
$(+)^{16 + d}(-)^{d}$ \cite{Nar}.
By varying the background fields one can deform all
(physically inequivalent) lattices of this type into each other.
The effect of such deformations can be made transparent by
chosing a special basis of $\Gamma_{16+d;d}$ \cite{Gin}. This basis
is constructed out of a basis ${\bf e}_{i}$ of the
compactification lattice $\Lambda$, a basis ${\bf e}^{*i}$
of its dual $\Lambda^{*}$ and a basis ${\bf e}_{A}$
of the even selfdual euclidean lattice $\Gamma_{16}$, which
contains the gauge quantum numbers of the corresponding
ten--dimensional theory:
\be     {\bf k}^{i} = \left( 0, \frac{1}{2} {\bf e}^{*i};
                      \frac{1}{2} {\bf e}^{*i} \right),
\label{basisvector_k}
\eq
\be     \ov{\bf k}_{i} = \left( {\bf A}_{i}, 2{\bf e}_{i} +
                D_{ij} \frac{1}{2} {\bf e}^{*j};
                D_{ij} \frac{1}{2} {\bf e}^{*j} \right),
\label{basisvector_kbar}
\eq
\be     {\bf l}_{A} = \left( {\bf e}_{A},
             -({\bf e}_{A} \cdot {\bf A}_{k}) \frac{1}{2} {\bf e}^{*k};
             -({\bf e}_{A} \cdot {\bf A}_{k}) \frac{1}{2} {\bf e}^{*k}
                \right),
\label{basisvector_l}
\eq
where
\be     D_{ij} = 2 \left( B_{ij} - G_{ij} - \frac{1}{4} ({\bf A}_{i}
                \cdot {\bf A}_{j} )  \right).
\label{D_matrix}
\eq
and $i,j,k =1,\ldots,d$, $A=1,\ldots,16$.
For all values of the background fields the only non--vanishing
(pseudo--euclidean) scalar products are
\be     {\bf k}^{i} \cdot \ov{\bf k}_{j} = \delta^{i}_{j}
        \mbox{  and  } {\bf l}_{A} \cdot {\bf l}_{B} =
        {\bf e}_{A} \cdot {\bf e}_{B} =: G_{AB}.
\eq
This basis is also useful for the definition of the twist.

\subsection{Orbifold compactifications}

The most general way to twist the Narain model is to
divide out some discrete group of automorphisms of the
Hilbert space \cite{NSV,GinRev}. This is usually done by specifying
automorphisms of the momentum lattice $\Gamma = \Gamma_{16+d;d}$
of the torus model \cite{NSV}.
We will restrict our attention to
\begin{itemize}
\item
cyclic groups of automorphims, which are generated by a
single twist $\Theta$ of finite order, $\Theta^{N} = {\bf 1}$,
\item
and to models, which admit an geometric interpretation as
compactification on a toroidal orbifold \cite{DHVW1,DHVW2}.
\end{itemize}
To fulfill the second condition, the lattice twist $\Theta$
must come from a {\em target space twist} $\theta \in O(d)$,
$\theta^{N} = {\bf 1}$, which defines the target space
${\bf O}^{d}$ of the $d$ internal coordinates,
\be     {\bf O}^{d} = {\bf T}^{d} / \la \theta \ra.
\eq
This is a well defined toroidal orbifold, provided
$\theta$ is a lattice automorphism of $\Lambda$.

Since the states $|({\bf K}, {\bf k}_{L}; {\bf k}_{R}) \ra$
are created by vertex operators
\be     \exp \left(
        K_{A} X_{L}^{A} (z) + k_{L,i} X^{i}_{L} (z) - k_{R, i} X^{i}_{R}
                (\ov{z})
        \right),
\eq
the choice of a target space twist $\theta$ acting on
$X^{i}(z, \ov{z}) $ $=$ $X^{i}_{L}(z)$ $+$ $X^{i}_{R}(\ov{z})$
does not only specify the action of the twist on the oszillators, but
also fixes the action of $\Theta$ on the $d + d$
quantum numbers $k_{L,i}, k_{R,i}$ of a state.
We are, however, free to define
the action of $\Theta$ on the 16 components $K_{A}$ by
specifying a {\em gauge twist}. It is necessary to choose a gauge twist
different from the identity, because otherwise a rank sixteen factor
of the gauge
group of the toroidal model remains unbroken. In general
the gauge twist can be an element of the sixteen dimensional
Euclidean group. The most studied possibility is that of a
pure shift (translation). This is technically the simplest
case, but it has two drawbacks \cite{IMNQ}:
\begin{itemize}
\item
The modding out by shifts leaves at least
one $U(1)$ gauge group per direction unbroken. Therefore
the rank of the gauge group cannot be reduced. In the case
of heterotic orbifolds the rank is at least 16, which is much
bigger than the rank of the gauge group of the standard model.
\item
Whereas some components of the metric and axionic background field
can still be varied continuously, the Wilson lines are
only allowed to take discrete values. Combined with point 1
this suggests that one
is only constructing special points of extended symmetry
in a bigger moduli space.
\end{itemize}
It is well known that if the gauge twist is a rotation (or contains
a rotation) the rank of the gauge group is generically reduced
and that some components of the Wilson lines can still be varied
continuously \cite{IMNQ}.
This has, however, not been studied systematically
enough to answer questions like these:
\begin{enumerate}
\item
What are the most general background fields compatible with
a given twist?
\item
What is the dimension of the moduli space? Which components
of the Wilson lines are moduli and which are discrete?
\item
How does the gauge group depend on the moduli? What patterns
of symmetry breaking do appear?
\end{enumerate}
These questions will be attacked in the following. For simplicity
the gauge twist will be a pure rotation,
$\theta' \in O(16)$, which must be an automorphism of $\Gamma_{16}$
in order to be compatible with the toric boundary conditions of
the sixteen extra left moving coordinates $X^{A}_{L}$.
Sometimes the gauge twist is defined in terms of the target
space twist. This imitates the embedding of the spin connection
into the gauge connection in Calabi--Yau
compactifications \cite{DHVW2}.
We will postpone
the specification of the gauge twist and try to learn first
to what extent it is restricted by simply demanding that
$\Theta$ is an lattice automorphism of $\Gamma_{16+d;d}$.
It will only be restricted
to have the same order as the target space twist,
$\theta'^{N} = {\bf 1}$, so that $\Theta$
has also order N.

In order to derive the consistency conditions between
the twist $\Theta$ and the background fields, it is convenient
to introduce the following matrices, which characterize the
action of the twist on the bases
${\bf e}_{i}$, ${\bf e}^{*i}$ and
${\bf e}_{A}$, $i = 1,\ldots,d$, $A=1,\ldots,16$ of the lattices
$\Lambda$, $\Lambda^{*}$ and $\Gamma_{16}$:
\be  \Theta: \; \left\{
\begin{array}{ccccc}
        {\bf e}_{i} & \rightarrow &\theta {\bf e}_{i} &
        = & \theta_{i}^{\;j} {\bf e}_{j}, \\
        {\bf e}^{*i} & \rightarrow &\theta {\bf e}^{*i} &
        = & \theta^{i}_{\;j} {\bf e}^{*j}, \\
        {\bf e}_{A} & \rightarrow & \theta'  {\bf e}_{A} &
        = & \theta_{A}^{\;B} {\bf e}_{B}. \\
        \end{array}
\label{twist_matrices} \right.
\eq
We can now start to analyze the {\em basic condition}
\be     \Theta \in \mbox{Aut}(\Gamma_{16+d;d})
\label{bascon}
\eq
in order to identify the admissible continuous and
discrete deformations of $\Gamma_{16+d;d}$.

\section{Derivation of the consistency equations}

The action of the twist on the momentum lattice is given by
(\ref{twist_matrices}). $\Theta$ is an automorphism of
$\Gamma_{16+d;d}$ if and only if the image of each basis vector
is again a lattice vector, that is an {\em integer} linear
combination of basis vectors. This will be worked out in the
following.

\subsection{The basis vectors ${\bf k}^{i}$ }

The first set of conditions is
\be      \Theta  {\bf k}^{i} = \left( 0,
         \frac{1}{2} \theta^{i}_{\;j} {\bf e}^{*j};
         \frac{1}{2} \theta^{i}_{\;j} {\bf e}^{*j}
         \right)
     \stackrel{!}{=}  M^{i}_{\;j} {\bf k}^{j} + M^{ij} \ov{\bf k}_{j}
                + M^{iA} {\bf l}_{A},
\eq
with  $M^{i}_{\;j}$, $M^{ij}$, $M^{iA}$ $\in \Zzahl$.
Comparing with  (\ref{basisvector_k}) - (\ref{basisvector_l})
immediately gives
\be  M^{i}_{\;j} = \theta^{i}_{j},\;\; M^{ij} = 0, \;\;M^{iA} =  0.
\label{T1}
\eq
The first of these equations means that $\theta$ must be
an automorphism (of $\Lambda^{*}$ and therefore)
of $\Lambda$. This condition is trivial from the
geometrical point of view, since the action of $\theta$ must
result in a well defined orbifold
${\bf O}^{6} = {\bf T}^{6} / \la \theta   \ra $.
It is however interesting that it follows automatically from the
basic condition (\ref{bascon}).
The other two equations simply tell us
that certain elements of the matrix of $\Theta$
(w.r.t. the basis (\ref{basisvector_k}) - (\ref{basisvector_l}) )
must vanish.

\subsection{The basis vectors $\ov{\bf k}_{i}$ }

The second set of conditions is less simple:
\be      \Theta  \ov{\bf k}_{i}  =  \left( \theta'
                {\bf A}_{i},  2 \theta_{i}^{\;j} {\bf e}_{j}  +
                \frac{1}{2} D_{ij} \theta^{j}_{\;k} {\bf e}^{*k};
                \frac{1}{2} D_{ij} \theta^{j}_{\;k} {\bf e}^{*k}
                \right)
         \stackrel{!}{=}  M_{ij} {\bf k}^{j}  +  M_{i}^{\;j}
                \ov{\bf k}_{j}  +  M_{i}^{\;A}  {\bf l}_{A},
\eq
with $M_{ij}$, $M_{i}^{\;j}$, $M_{i}^{\;A}$  $\in \Zzahl$.
We can now use again (\ref{basisvector_k}) - (\ref{basisvector_l}).
The resulting equation can be seperated into three equations, the
first one for the first sixteen components, the second
for the next six components and the third for the last
six components. Taking the difference between the second and third
equation yields
\be     2\theta_{i}^{\;j} {\bf e}_{j}  =  2 M_{i}^{\;j} {\bf e}_{j}.
\eq
Since the ${\bf e}_{i}$ are linearly independent, we get
\be     \theta_{i}^{\;j}  =  M_{i}^{\; j},
\label{T2}
\eq
which again states that $\theta \in \mbox{Aut}(\Lambda)$.
Using this, the first and third equation are
\be     \theta' {\bf A}_{i}  =  \theta_{i}^{\;j} {\bf A}_{j}
        + M_{i}^{\;A} {\bf e}_{A},
\label{generalized_embedding}
\eq
\be     D_{ij} \theta^{j}_{\;k}  -  \theta_{i}^{\;j} D_{jk}
        = M_{ik} - M_{i}^{\;A} ({\bf e}_{A} \cdot {\bf A}_{k} ).
\label{generalized_commutator}
\eq
Equation (\ref{generalized_embedding}) implies that, up to a lattice
vector of $\Gamma_{16}$, the Wilson lines must transform under
the gauge twist $\theta'$ like the directions to which they are
assigned to transform under the target space twist $\theta$.
This is a generalization of the usual construction of the gauge
twist by embedding the target space translation group
into the gauge group \cite{DHVW2}, where no lattice vector is allowed
($M_{i}^{\;A} = 0$).

Equation (\ref{generalized_commutator}) shows that the
D--matrix (\ref{D_matrix}) must commute with the twist
$\theta$, except for the right hand side. It resembles the equations
which give the number of moduli
in bosonic orbifolds and heterotic orbifolds without Wilson
lines \cite{EJL}.
Both equations will be analyzed in the next
section.

We close this subsection by giving formulas for
the matrix elements $M_{i}^{\;A}$, $M_{ik}$ of the
momentum lattice twist $\Theta$. This can be done by introducing
a matrix $G^{AB}$, which is the inverse of the lattice metric
\be   {\bf e}_{A} \cdot  {\bf e}_{B}   =  G_{AB}
\eq
of $\Gamma_{16}$. The result is:
\be     M_{i}^{\;A} = G^{AB} \left( (\theta' {\bf A}_{i} -
                \theta_{i}^{\;j} {\bf A}_{j}) \cdot {\bf e}_{B}
                \right) .
\label{T3}
\eq
\be   M_{ik}  = D_{ij} \theta^{j}_{\;k}  -  \theta_{i}^{\;j} D_{jk}
                + (\theta' {\bf A}_{i} - \theta_{i}^{\;j} {\bf A}_{j})
                \cdot {\bf A}_{k}.
\label{T4}
\eq

\subsection{The basis vectors ${\bf l}_{A}$ }

The third set of consistency conditions is
\be   \Theta {\bf l}_{A} = \left( \theta'  {\bf e}_{A},
        -\frac{1}{2} ({\bf e}_{A} \cdot {\bf A}_{i}) \theta^{i}_{\;j}
        {\bf e}^{*j};
        -\frac{1}{2} ({\bf e}_{A} \cdot {\bf A}_{i} )\theta^{i}_{\;j}
        {\bf e}^{*j}  \right)
  \stackrel{!}{=}  M_{Ai} {\bf k}^{i}  +  M_{A}^{\;i} \ov{\bf k}_{i}
        + M_{A}^{\;B} {\bf l}_{B},
\eq
with $M_{Ai}$, $M_{A}^{\;i}$, $M_{A}^{\;B}$ $\in \Zzahl$.
Applying the same procedure as in the last subsection
one derives
\be     M_{A}^{\;i} = 0.
\label{T5}
\eq
Substituting this in the equation for the first sixteen
components gives
\be     \theta' {\bf e}_{A}  = M_{A}^{\;B} {\bf e}_{B},
\eq
which means that the gauge twist is a lattice automorphism
of $\Gamma_{16}$. Again we have rederived a condition
which is sometimes stated independently from the basic condition
(\ref{bascon}).
Using (\ref{basisvector_k})
we have
\be    M_{A}^{\;B} = \theta_{A}^{\;B}.
\label{T6}
\eq
Using the previous results, the equation for the last six components
can be brought into the form
\be     {\bf e}_{A} \cdot  (\theta'^{-1} {\bf A}_{j} -
                {\bf A}_{i} \theta^{i}_{\;j} )  = M_{Aj}.
\label{T7}
\eq
This means that
$(\theta'^{-1} {\bf A}_{j} -$
${\bf A}_{i} \theta^{i}_{\;j} )$
must be in $\Gamma_{16}$, because this lattice is selfdual.
It is however no new condition, since it is equivalent to
(\ref{generalized_embedding}). (This is easily seen using
$\theta^{N} = {\bf 1}$, $\theta'^{N} = {\bf 1}$.)

Thus we achieved two results: The first is that the
basic condition $\Theta \in \mbox{Aut}(\Gamma_{16+d;d})$ is equivalent
to the equations (\ref{generalized_embedding}),
(\ref{generalized_commutator}) and $\theta$, $\theta'$
being automorphisms of $\Lambda$, $\Gamma_{16}$.
The second is an explicit matrix representation of
$\Theta$ w.r.t. the basis (\ref{basisvector_k}) - (\ref{basisvector_l})
in terms of the matrices of $\theta$, $\theta'$ and the
background fields. It is given by equations (\ref{T1}), (\ref{T2}),
(\ref{T3}), (\ref{T4}), (\ref{T5}), (\ref{T6}) and (\ref{T7}).

\section{Analysis of the consistency equations}

In this section the two consistency equations
(\ref{generalized_embedding}) and
(\ref{generalized_commutator}) will be analyzed.
One useful tool for dealing with such equations is to refer them
not to a lattice basis but to an orthonormal basis \cite{EJL}.
This is simple
linear algebra, but can easily cause confusion. Therefore
we will collect some useful formulas and introduce notation first.

\subsection{Lattice basis and orthonormal basis}
\label{LBOB}

In equations (\ref{generalized_embedding}) and
(\ref{generalized_commutator}) the target space twist
$\theta$, which is an orthogonal map on the real
vector space $\la \Lambda \ra_{\fs{R}}$ $\simeq$ $\Rzahl^{d}$,
appears via its matrices w.r.t. to the lattice bases of
$\Lambda$ and $\Lambda^{*}$:
\be     M_{ \{ {\bf e}_{i} \} }(\theta)
= \vartheta = (\theta_{i}^{\;j}),
        \;\;\;
        M_{ \{ {\bf e}^{*i} \} } (\theta)
        = \vartheta^{*} = (\theta^{i}_{\;j})
\eq
Since a generic lattice is neither orthogonal nor selfdual,
the matrix $\vartheta$ is in general not orthogonal and
not equal to $\vartheta^{*}$. But the orthogonality of the map
$\theta$ is reflected in the relation
\be     \vartheta^{*} = \vartheta^{T, -1}.
\eq
If we refer $\theta$ to an orthonormal basis ${\bf e}_{\mu}$,
\be     M_{ \{ {\bf e}_{\mu} \} } (\theta) = \ov{\vartheta}
        = (\theta_{\mu}^{\;\nu})
\eq
then the matrix $\ov{\vartheta}$ is orthogonal and
selfdual (w.r.t. our *--operation):
\be     \ov{\vartheta}\, \ov{\vartheta}^{T} = {\bf 1},\;\;\;
        \ov{\vartheta} = \ov{\vartheta}^{*}
\eq
This will be used to simplify equations  (\ref{generalized_embedding}),
(\ref{generalized_commutator}) later.

The two bases are connected by an invertible matrix ${\bf T}$,
defined by
\be     {\bf T} = (T_{i}^{\;\mu}),\;\;\;
        {\bf e}_{i} = T_{i}^{\;\mu} {\bf e}_{\mu}
\eq
The inverse matrix is denoted by ${\bf T}^{-1} = (T_{\mu}^{\;i})$.
The dual bases are connected by
\be     {\bf e}^{*i} = {\bf e}^{*\mu} T_{\mu}^{\;i}
\eq
and the relation between the various matrices represending
$\theta$ is
\be     \ov{\vartheta} = {\bf T}^{-1} \vartheta {\bf T}
        = {\bf T}^{T} \vartheta^{*} {\bf T}^{T, -1}
\eq
When dealing with equations (\ref{generalized_embedding}) and
(\ref{generalized_commutator}) we must also transform
the vectors ${\bf A}_{i}$ and
\be {\bf v}_{i} := M_{i}^{\;A}{\bf e}_{A}
\eq
 by
\be     {\bf A}_{\mu} = T_{\mu}^{\;i} {\bf A}_{i},\;\;\;
        {\bf v}_{\mu} = T_{\mu}^{\;i} {\bf v}_{i}.
\eq
This must not be confused with a basis transformation since
${\bf A}_{i}$ and ${\bf v}_{i}$ are not vectors in
$\la \Lambda \ra_{\fs{R}}$, but in a different,
sixteen dimensional vector space $\la \Gamma_{16} \ra_{\fs{R}}$
$\simeq \Rzahl^{16}$. We are simply using the invertible
matrix ${\bf T}$ to rearrange them into new linear
combinations\footnote{In the case of Wilson lines this can be
interpreted as another decomposition of the gauge connection:
${\bf A} = {\bf A}_{i} dx^{i}  = {\bf A}_{\mu} d x^{\mu}$.}.
Note also that the vectors ${\bf v}_{i}$ by definition
are lattice vectors of $\Gamma_{16}$ whereas the ${\bf v}_{\mu}$
generically are not. Likewise the matrices $\vartheta$,
$\vartheta^{*}$ are integer matrices, because they describe a
lattice automorphism w.r.t. a lattice basis, whereas
$\ov{\vartheta}$ will in general be no integer matrix.

Some final technical remarks: In the following component
and matrix notation as defined here will be freely interchanged.
Indices referring to the orthonormal basis will be written
lowercase, if no misinterpretation is possible (for example
$\theta_{\mu \nu} $$:= \theta_{\mu}^{\;\nu} $$ =  \theta^{\mu}_{\;\nu}$).

\subsection{Continuous and discrete Wilson lines}

The next step is to show that equation
(\ref{generalized_embedding}) decides which components
of the Wilson lines can be varied continuously in our
orbifold model and which components are restricted
to a discrete set of values. Referring to an
orthonormal basis, we get:
\be     \theta' {\bf A}_{\mu} = \theta_{\mu \nu} {\bf A}_{\nu}
                + {\bf v}_{\mu}.
\label{gen_emb}
\eq
In order to simplify this equation
we decompose each Wilson line in a part $\wt{\bf A}_{\mu}$
which is invariant under the gauge twist and its
complement $\wh{\bf A}_{\mu}$\footnote{Such a decomposition
is of cause also possible for the ${\bf A}_{i}$. We are,
however, preparing a second decomposition, which in general
forces us to switch to the ${\bf A}_{\mu}$.}.
This decomposition is both
orthogonal and direct (i.e. unique):
\be     {\bf A}_{\mu}  =  \wh{\bf A}_{\mu} +   \wt{\bf A}_{\mu},
        \;\;\;\theta'  \wt{\bf A}_{\mu}  =   \wt{\bf A}_{\mu}.
\eq
Doing the same with the inhomogeneous term
${\bf v}_{\mu}$ in
(\ref{gen_emb}) we get the two equations
\be     \theta'  \wt{\bf A}_{\mu} =  \wt{\bf A}_{\mu}  =
             \theta_{\mu \nu}
           \wt{\bf A}_{\nu}  +  \wt{\bf v}_{\mu},\;\;\;
       \theta'  \wh{\bf A}_{\mu} = \theta_{\mu \nu}
           \wh{\bf A}_{\nu}  +  \wh{\bf v}_{\mu}.
\label{invariant_noninvariant}
\eq

These equations can be decomposed further, if we seperate
such directions on ${\bf T}^{6}$  which are invariant
under the target space twist $\theta$ from those which are
non--invariant.
This is however (in general)
not possible w.r.t. the lattice basis ${\bf e}_{i}$ of
$\Lambda$\footnote{The author thanks D. Jungnickel for
pointing this out to him. See also \cite{ErlKle}
for implications on the classification of orbifolds.}.
But there always exists an orthonormal basis ${\bf e}_{\mu}$
such that the matrix of the twist is block--diagonal
\be      (\theta_{\mu \nu})  =  \left(  \begin{array}{cc}
                \theta_{ab}&  {\bf 0}  \\
                {\bf 0}  &  \delta_{mn}  \\
        \end{array} \right),
\label{tstwist_stbasis}
\eq
where $a,b = 1,\ldots,d-M$, $m,n = d-M+1,\ldots,d$. $M$ is the
dimension of the invariant subspace of $\theta$ and
the matrix $(\theta_{ab})$ does not have
the eigenvalue 1.
By (\ref{tstwist_stbasis}) the transformed equations
(\ref{invariant_noninvariant}) decouple into four equations.
There are four different
kinds of Wilson lines\footnote{Although $\wh{\bf A}_{a}$
and $\wt{\bf A}_{a}$ are strictly speaking different
projections of one Wilson line ${\bf A}_{a}$, we will
loosely refer to them as ``different Wilson lines'',
because they decouple and behave quite differently, as shown
in the text.},
each subject to a different constraint:
\begin{enumerate}
\item
Non--invariant Wilson lines, which are assigned to a
non--invarant direction in target space:
\be     \theta' \wh{\bf A}_{a} - \theta_{ab} \wh{\bf A}_{b}
         = \wh{\bf v}_{a}.
\label{wl1}
\eq
This is an inhomogeneous linear equation for the $\wh{\bf A}_{a}$.
Since the corresponding homogeneous equation has nontrivial
solutions\footnote{We could for example simply define
the gauge twist by $\theta' \wh{\bf A}_{a} $
$= \theta_{ab} \wh{\bf A}_{b}$.},
these Wilson lines can be varied continuously.
To calculate the number of moduli we must take into account
that they are further constrained by equation
(\ref{generalized_commutator}). This will be done in the next
section.
\item
Invariant Wilson lines assigned to non--invariant directions
in target space:
\be  \theta'\wt{\bf A}_{a} - \theta_{ab}\wt{\bf A}_{b} =
\wt{\bf A}_{a} - \theta_{ab}\wt{\bf A}_{b}
           = \wt{\bf v}_{a}.
\label{wl2}
\eq
Since the matrix $(\theta_{ab})$ has no eigenvalue
1, we could argue that the
the corresponding homogeneous equation has only
the trivial solution and the solution of the inhomogeneous
equation is unique. But this argument does not apply, when
the $\wt{\bf A}_{a}$ are not linear independent.
Fortunately we can modify a trick from
\cite{IMNQ}, which allows us to solve (\ref{wl2}) for the
Wilson lines. We can iteratively apply
$\theta'$ to equation (\ref{wl2}) and use the fact, that
$\wt{\bf A}_{a}$ and $\wt{\bf v}_{a}$ are invariant:
\be    \wt{\bf A}_{a} = \theta' \wt{\bf A}_{a} =
        \theta_{ab} \wt{\bf A}_{b} +  \wt{\bf v}_{a}
\eq
\be     \wt{\bf A}_{a} = \theta'^{2} \wt{\bf A}_{a}
        =\theta_{ab} \theta' \wt{\bf A}_{b}   +   \wt{\bf v}_{a}
        =
\eq
\[      = \theta_{ab} ( \theta_{bc} \wt{\bf A}_{c} + \wt{\bf v}_{b} )
                + \wt{\bf v}_{a}
        = \theta_{ab} \theta_{bc} \wt{\bf A}_{c}
                + \wt{\bf v}_{a} + \theta_{ab} {\bf v}_{b}
\]
and so on. Since $\theta^{N} = {\bf 1}$, the sum of the first
N powers of $(\theta_{ab})$ projects onto the invariant
subspace. But $(\theta_{ab})$ has no eigenvalue 1, implying
\be     \theta_{ab} + \theta_{ac} \theta_{cb}  +
        \cdots  \underbrace{\theta_{ac} \cdots \theta_{db} }
        _{\mbox{\small N factors}}
        = {\bf 0}.
\eq
Adding the first N equations produced by iteration and
using this identity, we arrive at an explicit expression
for $\wt{\bf A}_{a}$ in terms of $\wt{\bf v}_{a}$:
\be  \wt{\bf A}_{a} = \frac{N-1}{N} \wt{\bf v}_{a}   +
        \frac{N-2}{N} \theta_{ab} \wt{\bf v}_{b}   +
        \cdots   +
        \frac{1}{N} \underbrace{\theta_{ac} \cdots \theta_{db}}
        _{\mbox{\small N-2 factors}} \wt{\bf v}_{b}.
\label{ExpldWL}
\eq
Therefore the Wilson lines
$\wt{\bf A}_{a}$ can only take discrete values\footnote{
Applying the same procedure to the $\wh{\bf A}_{a}$ would lead
to the trivial identity $0=0$, giving no discretizing
constraint.}.
\item
Non--invariant Wilson lines assigned to invariant directions
in target space:
\be     \theta' \wh{\bf A}_{m}  -  \wh{\bf A}_{m}
                =  \wh{\bf v}_{m}.
\label{wl3}
\eq
The Wilson lines $\wh{\bf A}_{m}$ are by definition
orthogonal to the invariant subspace of the gauge twist
$\theta'$. The homogeneous equation has only the trivial
solution, the Wilson lines are discrete\footnote{In this
case the ``direct argument'' applies, because
$\theta' \wh{\bf A}_{m} = \wh{\bf A}_{m}$ implies
$\wh{\bf A}_{m} = 0$.}. A formula similar to
(\ref{ExpldWL}) can be derived by the iterative trick:
\be     \wh{\bf A}_{m} = - \frac{N-1}{N} \wh{\bf v}_{m} -
        \cdots - \frac{1}{N} \theta'^{N-2} \wh{\bf v}_{m}.
\eq
\item
Invariant Wilson lines assigned to invariant directions
in target space:
\be     \theta'  \wt{\bf A}_{m}  -  \wt{\bf A}_{m}
        =0        =  \wt{\bf v}_{m}.
\label{wl4}
\eq
This equation is identically fulfilled for
the Wilson lines $\wt{\bf A}_{m}$, which can be varied
continuously.
\end{enumerate}

Thus we have shown that equation (\ref{generalized_embedding})
can be used to disentangle the discrete and continouus
parts of the Wilson lines. The resulting picture is
simple and symmetric: (non) invariant Wilson lines
corresponding to (non) invariant directions are continuous,
whereas invariant (non--invariant) Wilson lines
corresponding ton non--invariant (invariant) directions
are discrete.

Our equations can be used to analyze
the role of discrete parameters further. Obviously there
are three kinds of discrete parameters connected with
the Wilson lines: Beside the discrete Wilson lines
$\wt{\bf A}_{a}$ and $\wh{\bf A}_{m}$ we must specify
a particular solution  $\wh{\bf A}^{\cal P}_{a}$
of the inhomogeneous equation (\ref{wl1}).
By equations (\ref{wl1}), (\ref{wl2}) and (\ref{wl3}) these discrete
Wilson parameters correspond to the
non--vanishing vectors $\wh{\bf v}_{a}$, $\wt{\bf v}_{a}$
and $\wh{\bf v}_{m}$. And these are related
to the matrix elements $M_{i}^{\;A}$ of the twist
$\Theta$. Furthermore, equation (\ref{generalized_embedding})
shows that our freedom in choosing $M_{i}^{\;A}$ and hence
discrete Wilson parameters, reflects our freedom in
defining the gauge twist $\theta'$. This shows
that discrete Wilson parameters should not be regarded as
moduli (which desribe continuous families of models with
the same twist) but as parts of the definition
of the gauge twist. This interpretation is familiar when
the gauge twist is realized by shifts:
In this case the Wilson lines are purely discrete, because N times
a Wilson  line must lie in the lattice $\Gamma_{16}$ and
their effect on the massless spectrum spectrum is to
impose extra projections, which eliminate all states that are
no ``Wilson singletts''. Thus adding discrete Wilson lines is
equivalent to modifying the twist by extra shift vectors \cite{INQ0}.

The allowed set of values for the discrete Wilson lines
in our construction is equal to that of the shift realization,
if we restrict our attention to
those twists, which admit the decomposition
\be      (\theta_{i}^{\;j})  =  \left(  \begin{array}{cc}
                \theta_{a}^{\;b} &  0  \\
                0  &  \delta_{m}^{\;n}  \\
        \end{array} \right),
\eq
w.r.t. a lattice basis. Geometrically, this is the case, when
the twist acts in such a way, that fixtori (if they appear)
factorize, such that the orbifold can be written as a
product of a torus and an orbifold without fixed tori.
(This is however not possible in general.)
Using also that $\Gamma_{16}$ is selfdual,
we can rewrite (\ref{generalized_embedding}) as
\be     {\bf e}_{A} \cdot (\theta' {\bf A}_{a} - \theta_{a}^{\;b}
                {\bf A}_{b} ) \in \Zzahl,\;\;\;
        {\bf e}_{A} \cdot (\theta' {\bf A}_{m} - \delta_{m}^{\;n}
                {\bf A}_{n}) \in \Zzahl,\;\;\;
        \forall A,a,m.
\label{gen_emb_var}
\eq
Using the fact that
\be     \Pi_{1} = \frac{1}{N} \sum_{N=1}^{N} \theta'^{n}
\eq
projects onto the invariant subspace and applying again the
iterative trick, we get
\be     {\bf e}_{A} \cdot N\wt{\bf A}_{a} \in \Zzahl,\;\;\;
        {\bf e}_{A} \cdot N\wh{\bf A}_{m} \in \Zzahl,\;\;\;
        \forall A,a,m.
\eq
Since $\wt{\bf A}_{a}$ is invariant under $\theta'$, and
$\wh{\bf A}_{m}$ orthogonal to the invariant subspace of
$\theta'$ this is equivalent to
\be     N \wt{\bf A}_{a} \in I_{\Gamma_{16}}
        := \{ {\bf w} \in \Gamma_{16} | \theta' {\bf w} = {\bf w}
                \},\;\;\; \forall a,
\eq
and
\be     N \wh{\bf A}_{m} \in N_{\Gamma_{16}}
        := \{ {\bf w} \in \Gamma_{16} | {\bf w} \cdot I_{\Gamma_{16}}
                = 0  \},\;\;\; \forall m,
\eq
where $I_{\Gamma_{16}}$ is the non--invariant sublattice of
$\Gamma_{16}$ and $N_{\Gamma_{16}}$ its orthogonal complement.
These constraints for discrete Wilson lines resemble those
for the purely discrete Wilson lines in the case of
shift realization of the gauge twist \cite{DHVW2}. This adds further
evidence to the claim that this construction yields only special
points in the complete heterotic orbifold moduli space.

\subsection{From the D--matrix to the number of moduli (Part I)}

In this section we will start to derive the number of moduli, which is
(almost) given by equation (\ref{generalized_commutator}).
After plugging (\ref{generalized_embedding}) into
(\ref{generalized_commutator}), we get
\be     D_{ij}\theta^{j}_{\;k}  - \theta_{i}^{\;j} D_{jk} = M_{ik}
        -  (\theta' {\bf A}_{i} - \theta_{i}^{\;j} {\bf A}_{j})
        \cdot {\bf A}_{k}.
\eq
Our goal is to solve this for the matrix ${\bf D} = (D_{ij})$.
The first
step is to convert the left side of the equation into a
proper commutator of matrices. This can be done by
referring to an orthonormal basis instead of a lattice basis.
This basis can of cause be chosen in a way that
the decomposition (\ref{tstwist_stbasis}) holds.
The equation is now
\be     D_{\mu \rho}\theta_{\rho \nu}  - \theta_{\mu \rho} D_{\rho \nu}
        = M_{\mu \nu}
        -  (\theta' {\bf A}_{\mu} - \theta_{\mu \rho} {\bf A}_{\rho})
        \cdot {\bf A}_{\nu},
\label{gen_com}
\eq
with\footnote{The definitions of ${\bf A}_{\mu}$
and $\theta_{\mu \nu}$ were given in section \ref{LBOB}. }
\be     D_{\mu \nu} = T_{\mu}^{\;i} D_{ij} T^{j}_{\;\nu},\;\;
        M_{\mu \nu} = T_{\mu}^{\;i} M_{ij} T^{j}_{\;\nu},\;\;
\eq
where, in terms of the matrix ${\bf T} = (T_{i}^{\;\mu})$
introduced in section \ref{LBOB}:
\be     (T_{\mu}^{\;i}) = {\bf T}^{-1},\;\;\;
        (T^{i}_{\;\mu}) = {\bf T}^{T, -1}.
\eq

The second step is to get rid of the second, Wilson line
dependent term on the right hand side. Using (\ref{tstwist_stbasis})
we observe that the discrete Wilson lines can be absorbed by
a redefinition of the matrix $M_{\mu \nu}$, whereas the
continuous Wilson lines can be absorped by the matrix $D_{\mu \nu}$:
\be        D_{ac}\theta_{cb} - \theta_{ac} D_{cb} =
        \underbrace{M_{ab}
        -  (\wt{\bf A}_{a}  -  \theta_{ac}\wt{\bf A}_{c}) \cdot
        \wt{\bf A}_{b} }_{M'_{ab}}.
\eq
\be     D_{mn}\delta_{np}  - \delta_{mn}D_{np} = 0
        = \underbrace{
        M_{mp} - (\theta' \wh{\bf A}_{m} - \wh{\bf A}_{m}) \cdot
        \wh{\bf A}_{p}}_{M'_{mp}}.
\eq
\be      \underbrace{(D_{am}
              + \wt{\bf A}_{a} \cdot \wt{\bf A}_{m})}_{D'_{am}}
              \delta_{mn}
     - \theta_{ab}
         \underbrace{ ( D_{bn}  +  \wt{\bf A}_{b} \cdot  \wt{\bf A}_{n})}
        _{D'_{bn}}
        = M_{an}.
\eq
\be  \underbrace{ ( D_{mc} + (\wh{\bf A}_{m} \cdot \wh{\bf A}_{c}) )}
        _{D'_{mc}} \theta_{cb}    -
        \delta_{mn} \underbrace{ (D_{nb} +
                (\wh{\bf A}_{n} \cdot \wh{\bf A}_{b}) )}_{D'_{nb}}
        = M_{mb}.
\eq
Summarizing in matrix notation, with $\ov{\bf D}' = (D'_{\mu \nu})$
and $\ov{\bf M}' = (M'_{\mu \nu})$, we have transformed
(\ref{generalized_commutator}) into the inhomogeneous
linear matrix equation
\be     [\ov{\bf D}', \ov{\vartheta} ] = \ov{\bf M}'.
\eq
To solve it, we need a particular solution of the inhomogeneous
equation and the general solution of the corresponding
homogeneous equation. This problem has been solved in
reference \cite{EJL} for purely symmetric and purely
antisymmetric matrices\footnote{In \cite{EJL}
consistency conditions for bosonic orbifolds were analyzed, with
the result, that deformations of the compactification lattice,
parametrized by a symmetric matrix ${\bf S}$, and the
background field, described by an antisymmetric matrix
${\bf B}$
must commute with the target space twist.
In case of the antisymmetric background field this must only
hold
up to a constant matrix.}.
The formula found for the particular solution also
applies to the general case. The resulting
matrix $\ov{\bf D}^{\cal P}$ consists of discrete parameters
which reflect our freedom of chosing the matrix elements
$M_{ik}$ of the twist $\Theta$. Like the discrete
Wilson parameters discussed in the last section, they can be
regarded as part of the definition of the twist. We will not
analyze them further, but instead focus on the moduli.

The general solution
of the homogeneous equation
\be     [\ov{\bf D}', \ov{\vartheta} ]  = 0
\label{hom_com}
\eq
can be found by generalizing the calculation of \cite{EJL}
to the case of an arbitrary matrix:
Remembering that the matrix $\ov{\vartheta}$ of the
target space twist $\theta$ w.r.t. an orthonormal basis
is an orthogonal matrix with $\ov{\vartheta}^{N}= {\bf 1}$
we can transform it by an orthogonal transformation
to its standard form
\be     \ov{\vartheta} = \left( \begin{array}{ccc}
         \tau & {\bf 0} & {\bf 0} \\
        {\bf 0} & -{\bf 1}_{l} & {\bf 0} \\
        {\bf 0} & 0& {\bf 1}_{M} \\
        \end{array}  \right),
\eq
where $\tau$ is an orthogonal $2n \times 2n$ matrix,
$2n + l + M = d$, and its eigenvalues are N--th root
of unity, but not $\pm 1$. The matrix $\tau$ can
be brought to the standard form
\be     \tau = \left(  \begin{array}{ccc}
        \tau_{1} & \cdots  & {\bf 0} \\
        \vdots & \ddots & \vdots \\
        {\bf 0} & \cdots  & \tau_{n} \\
        \end{array}  \right).
\eq
where the $2 \times 2$ blocks $\tau_{\alpha}$ correspond to
complex conjugated pairs of eigenvalues
$\exp(\pm 2\pi ik_{\alpha})$,
$0<k_{\alpha}<1$, $k_{\alpha} \not=\frac{1}{2}$, $Nk_{\alpha} = 1$,
$\alpha=1,\ldots,n$:
\be     \tau_{\alpha}  =  \left( \begin{array}{cc}
        c_{\alpha} & -s_{\alpha} \\ s_{\alpha} & c_{\alpha} \\
        \end{array}  \right) , \;\;\;c_{\alpha} =
        \cos(2 \pi  k_{\alpha}),
        \;s_{\alpha} = \sin(2 \pi  k_{\alpha}).
\eq
Plugging this special form of $\ov{\vartheta}$ into (\ref{hom_com})
yields
\be     \ov{\bf D}' =  \left( \begin{array}{ccc}
                      {\bf d} & {\bf 0} & {\bf 0} \\
                     {\bf 0} & {\bf p} & {\bf 0} \\
                     {\bf 0} & {\bf 0} & {\bf q} \\
                     \end{array}  \right).
\label{D_block}
\eq
The $l \times l$ matrix ${\bf p}$ and the $M \times M$ matrix
${\bf q}$ are arbitrary, whereas the $2n \times 2n$ matrix
${\bf d}$ must commute with $\tau$:
\be     [{\bf d}, \tau]  = {\bf 0}.
\eq
Decomposing ${\bf d}$ into $n^{2}$ $2 \times 2$ blocks
${\bf D}_{\alpha \beta}$, $\alpha, \beta = 1,\ldots,n$, the equation
reduces to
\be     {\bf D}_{\alpha \beta} \tau_{\beta}  - \tau_{\alpha}
          {\bf D}_{\alpha \beta}  = 0 \;\;\; \forall \alpha,\beta.
\label{block_com}
\eq
If we set for arbitrary but fixed $\alpha, \beta$
\be     {\bf D}_{\alpha \beta} = \left( \begin{array}{cc}
                p &q \\  r & t \\  \end{array}  \right),
\eq
we get a set of four homogeneous, linear equations in four
real variables $p, q, r, t$. The necessary and sufficent condition
for the existence of a nontrivial solution is the vanishing of
the characteristic determinant of that system. Using identities
for trigonometric functions this reduces to
\be     c_{\alpha} = c_{\beta},
\eq
which implies
\be     s_{\alpha} = \pm s_{\beta}.
\eq
Therefore equation (\ref{block_com}) has two types of solutions:
\begin{itemize}
\item
If
$s_{\alpha} = s_{\beta}$, which is equivalent to
$k_{\alpha} = k_{\beta}$:
\be     q = -r,\;\;\;p=t.
\eq
\item
If
$s_{\alpha} = -s_{\beta}$, which is equivalent to
$k_{\beta} = 1- k_{\alpha}$:
\be     q = r,\;\;\;p=-t.
\eq
\end{itemize}
Thus each nonvanishing ${\bf D}_{\alpha \beta}$ yields two continuous
parameters, whereas the unconstrained matrices ${\bf p}$, ${\bf q}$
give another $M^{2} + l^{2}$ parameters.
Denoting the dimension  of the eigenspace to
eigenvalue $\exp(2 \pi i k)$ by $d_{k}$,
the number of independent solutions of the homogeneous equation
(\ref{hom_com}) is
\be     M_{D} = d_{1}^{2} + d_{\frac{1}{2}}^{2}
        + 2 \sum_{k \in \fs{Q} | 0< k< \frac{1}{2}  } d_{k}^{2}
        = \sum_{k \in \fs{Q} | 0 \leq k <1 } d_{k}^{2}.
\label{D_moduli}
\eq
This formula looks like a generalization of the formulas
given in \cite{EJL} for the
metric and axionic moduli of bosonic orbifolds.
However, it does not give the number of moduli directly.
In order to identify the moduli in the heterotic case,
we have to decompose the matrix $D'_{\mu \nu}$ into the
background fields. But if we want to give explicit
formulas for the Wilson moduli, we must specify the gauge twist
first.

\subsection{Definition of the gauge twist}

According to equation $(\ref{generalized_commutator})$ the
Wilson lines must transform up to lattice vectors of $\Gamma_{16}$
under the gauge twist
as the corresponding directions of target space
transform under the target space twist. This resembles the
embedding of the spin connection into the gauge connection
used in Calabi--Yau compactifications and suggests that
``somehow'' the gauge twist is ``identical'' to the
target space twist. This can be made precise in two steps\footnote{
We will see later, that this construction, together with a
modified version, gives in the case of $\Zzahl_{3}$ orbifolds
based on $E_{8} \otimes E_{8}$ models all modular invariant
theories. Thus it seems not to be very restrictive.}:
First we define the gauge twist by {\em lifting} the
target space twist. To do this, we select a
$d$--dimensional
sublattice $\Lambda'$ of $\Gamma_{16}$ with lattice basis
$\wh{\bf e}_{i}$, such that a bijective map (isomorphism of groups)
\be   \phi: \; \Lambda \rightarrow \Lambda' \subset \Gamma_{16}:
        \; {\bf e}_{i} \rightarrow \wh{\bf e}_{i}
\eq
exists.
Then we decompose the sixteen dimensional
space spanned by $\Gamma_{16}$
\be    \la \Gamma_{16} \ra_{\fs{R}} =
        \la \Lambda' \ra_{\fs{R}}  \oplus
        \la \Lambda' \ra_{\fs{R}}^{\bot}
\eq
and define the gauge twist by
\be     \theta' = \phi \theta \phi^{-1}  \oplus {\bf 1}.
\eq
The second step is a variant of {\em embedding} the target space
translation group into the gauge twist group \cite{DHVW2}.
This is
usually done by a map between the lattice basis of $\Lambda$
and the Wilson lines,
\be     \chi: \; \Lambda \rightarrow  \la \Gamma_{16} \ra_{\fs{R}}:
        \; {\bf e}_{i} \rightarrow {\bf A}_{i}
\label{chi}
\eq
which must commute with the twist:
\be     \theta' \chi = \chi \theta
\eq
This implies
\be     {\bf A}_{i} = \theta_{i}^{\;j} {\bf A}_{j}
\eq
which is equivalent to ${\bf v}_{i} = 0$ or $M_{i}^{\;A}=0$,
implying that the Wilson lines must strictly transform like
the corresponding directions in target space. This is, however,
to restrictive in the context of our construction, because it
does not allow for discrete Wilson lines ($\wt{\bf v}_{a} = {\bf 0}$
and $\wh{\bf v}_{m} = {\bf 0})$, which are, from our point of
view, parts of the definition of the twist, and no (continuous)
background fields. Therefore, we will only apply the milder
constraint $\wh{\bf v}_{a} = {\bf 0}$: This allows discrete
Wilson lines $\wt{\bf A}_{a}$, $\wh{\bf A}_{m}$, whereas
the continuous Wilson lines transform strictly like the
corresponding directions:
\be     \theta' \wh{\bf A}_{a} = \theta_{ab} \wh{\bf A}_{b},\;\;\;
        \theta' \wh{\bf A}_{m} = \delta_{mn} \wh{\bf A}_{n} =
        \wh{\bf A}_{m}.
\eq
Thus our modification amounts to applying the embedding $(\ref{chi})$
only to
the continuous (parts of the) Wilson lines.

\subsection{From the D--matrix to the number of moduli (Part II)}

We are now ready to calculate the number of moduli. First let us
decompose the matrix $\ov{\bf D} = (D_{\mu \nu})$ into the
background fields:
\be     D_{\mu \nu} =
         2 \left(  B_{\mu \nu}  -  G_{\mu \nu}
        - \frac{1}{4}
        ({\bf A}_{\mu}  \cdot {\bf A}_{\nu})  \right).
\eq
Note that the metric moduli, which describe the allowed
deformations of $\Lambda$ and can be parametrized by the
lattice metric $G_{ij}$ do not appear in the D--matrix, when
it is refered to an orthonormal basis, since
\be     G_{\mu \nu} = e_{\mu} \cdot e_{\nu} = \delta_{\mu \nu}.
\eq
They can however be calculated from the constraint
$\theta \in \mbox{Aut}(\Lambda)$, as will be shown below.
Now, the antisymmetric and symmetric part of $\ov{\bf D}$
correspond to the axionic moduli and to some Wilson moduli,
parametrizing those deformations of the Wilson lines
which change the matrix of scalar products
$({\bf A}_{\mu} \cdot {\bf A}_{\nu})$ and will therefore be called
{\em metric Wilson moduli}. There will also be {\em isometric
Wilson moduli} describing orthogonal transformations,
because rotations of the Wilson lines relative to the basis
${\bf e}_{A}$ of $\Gamma_{16}$ change the lattice $\Gamma_{16+d;d}$
and therefore the physical spectrum.

To get the desired formulas, let us first assume that the
target space twist has no fixtori ($d_{1} = 0$).
Then the allowed Wilson lines are $\wh{\bf A}_{a}$,
$\wt{\bf A}_{a}$, $a=1,\ldots,d$, where the $\wh{\bf A}_{a}$
take continuous values in the $d$--dimensional subspace
$\la \Lambda' \ra_{\fs{R}}$ and
the $\wt{\bf A}_{a}$ take discrete values in
$\la \Lambda' \ra_{\fs{R}}^{\bot}$. In this case there is
no difference between the original matrix $(D_{ab})$ and
the redefined matrix $(D_{ab}')$, and (\ref{D_moduli}) gives
the sum of the number $M_{B}$
of axionic moduli and the number $M_{A,m}$ of metric Wilson moduli.
In order to separate them, we have to impose the extra
constraint of antisymmetry or symmetry on the D--matrix.
The result is
(see also \cite{EJL}) that both the antisymmetric and the symmetric
part contain half of the moduli corresponding to the complex
eigenvalues. As the part correponding to the eigenvalues
$-1$ (the matrix ${\bf p}$ in (\ref{D_block})) is unconstrained, we get:
\be     M_{B}  =
        \left(
        \begin{array}{c} d_{\frac{1}{2}} \\ 2 \\
        \end{array} \right)
        + \sum_{k \in \fs{Q}|0<k<\frac{1}{2}}
        d_{k}^{2}.
\eq
\be      M_{A,m} =
         \left(
        \begin{array}{c}  d_{\frac{1}{2}} +1 \\
        2 \\  \end{array} \right)
        +\sum_{k \in \fs{Q}|0<k<\frac{1}{2}}
        d_{k}^{2}.
\eq
The number of axionic moduli is the same as in the bosonic case
\cite{EJL}.

The number $M_{G}$ of metric moduli can now be extracted from the
condition $\theta \in \mbox{Aut}(\Lambda)$ in the same way as in bosonic
orbifold models \cite{EJL}.
Let $\sigma$ be an invertible map deforming
$\Lambda$ into another lattice $\Lambda'$:
\be     \sigma: \; \Lambda \rightarrow \Lambda':
        \; {\bf e}_{i} \rightarrow {\bf e}'_{i}
\eq
This mapping must not be confused with the coordinate
transformation  ${\bf e}_{i} \rightarrow {\bf e}_{\mu}$
introduced earlier.
A deformation of $\Lambda$ is compatible with a given
target space twist, if and only if $\theta$ is also an
automorphism of $\Lambda'$. Therefore the deformation must
commute with the twist:
\be     \sigma \theta = \theta \sigma
\eq
Since the lattice  $\Lambda$ enters physics
only through scalar products,
only those deformations changing the lattice metric are
moduli:
\be   G_{ij} = {\bf e}_{i} \cdot {\bf e}_{j} \rightarrow
      G'_{ij} = {\bf e}'_{i} \cdot {\bf e}'_{j}
\eq
This implies that we must consider symmetric maps only\footnote{
Remember that an invertible map can be decomposed into
an orthogonal and a symmetric map by a polar decomposition.}.
Referring
to an orthonormal basis, such a map can be represented by
a symmetric matrix $\ov{\bf S}$. The resulting matrix equation
\be     [\ov{\bf S}, \ov{\vartheta}] = 0
\eq
is formally equivalent to that for the metric Wilson moduli, implying
\be     M_{G} = M_{A,m}.
\eq

The number $M_{A,i}$ of isometric Wilson moduli can be calculated
similarly\footnote{The argument given here in the first version
of the paper misses the point under consideration. The author
thanks Gabriel Lopes Cardoso for pointing out that the total
number of Wilson moduli was to small in the case of degenerate
eigenvalues of the twist.}.
Let $\rho$ be an orthogonal transformation acting non--trivially
on the non--invariant
subspace of $\la \Gamma_{16} \ra_{\fs{R}}$
and as the identity on the invariant subspace.
Let ${\bf A}_{a}$ be Wilson lines which
solve equations (\ref{gen_emb}) and
(\ref{gen_com}).
Since the target space twist has by assumption no
fixtori, all six Wilson lines can be varied continuously on
the non--invariant subspace
$\la \Lambda' \ra_{\fs{R}}$ and are discrete on the complement.
If we deform the continuous parts $\wh{\bf A}_{a}$
of the Wilson lines by $\rho$,
then the new Wilson lines
\footnote{Note that
rotations of the Wilson lines relative to $\Gamma_{16}$ change
the scalar products ${\bf A}_{a} \cdot {\bf e}_{A}$ and are
therefore physically relevant.}
$\rho {\bf A}_{a}$ are automatically solutions of
(\ref{gen_com}), because $\rho$ is orthogonal.
And they are solutions of (\ref{gen_emb}) if and only if
the deformation $\rho$ commutes with the gauge twist
$\theta'$:
\be     \theta' \rho = \rho \; \theta'.
\label{isomet}
\eq
Since $\rho$ is a continuous deformation of the old Wilson lines
and therefore continuously connected to the identity, its matrix
w.r.t. a orthonormal frame is in $SO(d)$ and can be written as the
exponential of an antisymmetric matrix:
\be  \ov{R} = \exp(a),\;\; a = -a^{t} \in so(d).
\eq
Thus equation (\ref{isomet}) is formally equivallent to the
equation which gives the number of axionic moduli:
\be   [\ov{\vartheta}', \ov{R}] = 0 \Leftrightarrow
      [\ov{\vartheta}',a ] = 0.
\eq
This implies that the number of isometric Wilson moduli equals
the number of axionic moduli:
\be    M_{A,i} = M_{B}.
\eq
The total number of Wilson moduli in the case without fixtori
is:
\be  M_{A} = M_{A,m} + M_{A,i} = M_{G} + M_{B} = M_{D}.
\eq

Let us now study what changes, when the target space twist
has a $M$--dimensional invariant subspace ($M < d$, $d_{1} = M$).
Decomposing it as
\be     \theta = \wh{\theta} \oplus {\bf 1}_{M}
\eq
we observe that the invariant subspace of the gauge twist
also grows:
\be     \theta' = \phi \wh{\theta} \phi^{-1} \oplus {\bf 1}_{M}
                \oplus {\bf 1}_{16-d}  =
                \phi \wh{\theta} \phi^{-1} \oplus
                {\bf 1}_{16-d+M}
\eq
There are two kinds of continuous Wilson lines:
$d - M$ Wilson lines $\wh{\bf A}_{a}$ taking values in
the $d-M$ dimensional non--invariant subspace $\wh{V}$ of
$\la \Gamma_{16} \ra_{\fs{R}}$ and $M$ Wilson lines
$\wt{\bf A}_{m}$ taking values in the $16-d+M$ dimensional
invariant subspace $\wt{V}$. As there are also discrete Wilson lines
$\wt{\bf A}_{a}$, $\wh{\bf A}_{m}$, we must take into account
the additional off diagonal blocks which distinguish
the modified matrix $D_{\mu \nu}'$ from $D_{\mu \nu}$.
By (\ref{D_block}) the equations for the off diagonal blocks are
\be     D'_{am} = -2 B_{am} - \frac{1}{2}
        \wt{\bf A}_{a} \cdot \wt{\bf A}_{m} = 0
\eq
and
\be     D'_{ma} = -2 B_{ma} - \frac{1}{2}
        \wh{\bf A}_{m} \cdot \wh{\bf A}_{a} = 0.
\eq
Using the fact that scalar products are symmetric and that the
B--matrix is antisymmetric, adding both equations gives:
\be     \wt{\bf A}_{a} \cdot \wt{\bf A}_{m} +
        \wh{\bf A}_{a} \cdot \wh{\bf A}_{m} =
        {\bf A}_{a} \cdot {\bf A}_{m} = 0
\label{A_tilde_A_hat}
\eq
which means that Wilson lines assigned to invariant and those
assigned to non--invariant directions must be pairwise
orthogonal. Subtracting the equations for the off diagonal blocks
yields
\be     8  B_{am} - \wt{\bf A}_{a} \cdot \wt{\bf A}_{m}
        + \wh{\bf A}_{a} \cdot \wh{\bf A}_{m} = 0
\eq
Eliminating the invariant parts of the Wilson lines we get
\be     4 B_{am} = -4 B_{ma} = - \wh{\bf A}_{a} \cdot \wh{\bf A}_{m}
\label{B_A_hat}
\eq
This shows that we can arbitrarily vary the continuous Wilson lines
$\wh{\bf A}_{a}$, as long as we tune the B--matrix according
to (\ref{B_A_hat})
and restrict the continuous Wilson lines $\wt{\bf A}_{m}$ such
that (\ref{A_tilde_A_hat}) holds.
This prescription gives $F\leq$$M(d-M)$ constraints for the
$\wt{\bf A}_{m}$, the precise number depending on how many
linear independent (especially non--vanishing) discrete Wilson
lines are chosen.
The $\wt{\bf A}_{a}$
remain unconstrained. This parametrization of the constraints and
moduli associated with the off--diagonal blocks of the
D--matrix will turn out to be convenient.

The analysis of the metric and isometric moduli associated
with the $\wh{\bf A}_{a}$ runs now parallel to the discussion
of continuous Wilson lines in models without fixtori, because
the diagonal block $D_{ab}$ is not modified (and because there
is no extra constraint). The result is
\be     \wh{M}_{A} = d_{\frac{1}{2}}^{2} +
         2 \sum_{k \in \fs{Q}|0<k<\frac{1}{2}} d_{k}^{2}.
\eq
Looking at the block $D_{mn}$ we see that for the metric
Wilson moduli of the $\wt{\bf A}_{m}$ no constraints arise.
As the twist acts as the identity in this subspace,
any deformation of the continuous Wilson lines commutes with it.
Thus the only restriction is the one resulting from
the off diagonal blocks of $D'_{\mu \nu}$:
\be     \wt{M}_{A} = M (16 -d + M)  - F
        \geq
        M (16 -d + M) - M (d - M)
        = M ( 16 -2d + 2M).
\eq

The modifications of the formulas for the metric and axionic
moduli are derived by observing that there are no constraints
for the ${\bf q}$ block in (\ref{D_block}). Taking into account that
metric (axionic) deformations are symmetric (antisymmetric)
we get:
\be      M_{G} =
         \left(
        \begin{array}{c}  d_{1} +1 \\
        2 \\  \end{array} \right)  +
         \left(
        \begin{array}{c}  d_{\frac{1}{2}} +1 \\
        2 \\  \end{array} \right)
        +\sum_{k \in \fs{Q}|0<k<\frac{1}{2}}
        d_{k}^{2}.
\eq
\be     M_{B}  =
         \left(
        \begin{array}{c}  d_{1} \\
        2 \\  \end{array} \right) +
        \left(
        \begin{array}{c} d_{\frac{1}{2}} \\ 2 \\
        \end{array} \right)
        + \sum_{k \in \fs{Q}|0<k<\frac{1}{2}}
        d_{k}^{2}.
\eq
This completes the calculation of the number of moduli
for all models with gauge twist defined by lifting of the
target space twist.

\subsection{Multiple liftings}

A simple generalization of the definition of the gauge twist
by lifting the target space twist is provided by liftings
to different, orthogonal sublattices $\Lambda'$, $\Lambda''$,\ldots
of $\Gamma_{16}$. If one compactifies six dimensions, there is,
for dimensional reasons, only the possibility of simple and
double liftings. In the second case there are two group isomorphisms
\be     \phi': \Lambda \rightarrow \Lambda' \subset \Gamma_{16},
        \;\;\;
        \phi'': \Lambda \rightarrow \Lambda'' \subset \Gamma_{16},
        \;\;\;
        \Lambda' \bot \, \Lambda'',
\eq
and the gauge twist is
\be       \theta' = \phi'\theta \phi'^{-1} \oplus
                \phi''\theta \phi''^{-1} \oplus
                {\bf 1}_{4}.
\eq
The generalization of the formulas derived for simple liftings
should be straightforward. For example it is clear, that the
number of moduli coming from continuous $\wh{\bf A}_{a}$
Wilson lines is doubled.

\section{The standard $\Zzahl_{3}$ orbifold with continuous
Wilson lines}

\subsection{Gauge twists for $\Zzahl_{3}$ orbifolds}

The construction of $\Zzahl_{3}$ orbifolds \cite{DHVW1}
is based on
toroidal compactifications with the compactification lattice
$\Lambda$ proportional to three copies of the root lattice
of the Lie algebra $A_{2}$ (which is the complex form of $su(3)$):
\be  \Lambda \propto A_{2}^{\oplus 3} = A_{2} \oplus A_{2} \oplus
        A_{2}
\eq
(We denote the root lattice with the same symbol as the complex Lie
algebra.)
The twist acts by simultanously rotating with an angle of 120 degree in
all three $A_{2}$ lattices. This is the Coxeter twist\footnote{
The Coxeter twist is the product of all fundamental Weyl reflections.
For background material on Lie algebras see for example \cite{Cor}.}
of the
corresponding Lie algebra. According to our formulas, there are
9 metric and 9 axionic moduli. The metric moduli correspond
to three radii and six angles, since we can scale the three
pairs of simple roots of the $A_{2}$ lattices and are also
free to rotate the three planes spanned by them relative
to each other, whereas the angle between the two simple
roots of each $A_{2}$ is fixed.

The most basic constraint on the gauge twist is that it must be
an autormophism of order three
of the lattice $\Gamma_{16}$, which is
chosen to be the $E_{8} \oplus E_{8}$ root lattice:
\be     \theta' \in \mbox{Aut}(E_{8} \oplus E_{8}),
        \;\;\;
        \theta'^{3} = {\bf 1}.
\eq
The Lie algebra $E_{8}$ has, up to conjugation, 112 automorphisms
of finite order,
which are listed in \cite{HolMyh}. They are all inner automorphisms
and therefore induced by Weyl reflections of the root lattice.
Among these, there are precisely four automorphisms of
order three. These are induced by simultanous
Coxeter twists in one, two,
three or four pairwise orthogonal $A_{2}$ sublattices.
Therefore, $E_{8} \oplus E_{8}$ has sixteen Weyl automorphisms of order
three. However, modular invariance leads to constraints on the
eigenvalues of the twist and only allows simultanous rotations
in three or in six $A_{2}$ sublattices of the $E_{8} \oplus E_{8}$
lattice \cite{NSV,INQ0}.
Note that these cases singled out by modular invariance
correspond precisely to simple or double liftings
of the target space twist. Whether this observation is a pure
coincidence or implies that the lifting construction of gauge
twists is preferred is far from obvious. By different
choices of the lift, there are altogether four inequivalent
$\Zzahl_{3}$ orbifold compactifications of the ten--dimensional
$E_{8} \oplus E_{8}$ string theory \cite{INQ0}. The standard version is
given by twisting
a $A_{2}^{\oplus 3}$ sublattice of one
$E_{8}$. This is the singular limit of a Calabi--Yau compactification
with identification of the holonomy group $SU(3)$
with the $SU(3)$ subgroup of one $E(8)$ group \cite{DHVW1,GSW}.
The unbroken gauge Lie algebra of the orbifold
model without Wilson lines is then $E_{6} \oplus A_{2} \oplus E_{8}$.
In the following we will study the effect of continuous
and discrete Wilson lines on the Lie algebra.
The first step is to introduce a convenient parametrization of
the $E_{8}$ lattice, the gauge twist and the Wilson lines.

\subsection{Parametrization of the $E_{8}$ lattice}

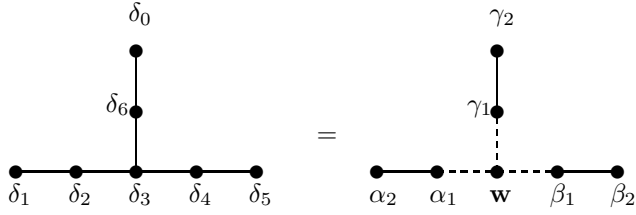
\begin{figure}
\unitlength0.8cm
\begin{center}
\begin{picture}(14,3)
\put(0,1){\line(1,0){4}}
\put(2,1){\line(0,1){2}}
\multiput(0,1)(1,0){5}{\circle*{0.2}}
\put(2,2){\circle*{0.2}}
\put(2,3){\circle*{0.2}}
\put(-0.125,0.5){$\delta_{1}$}
\put(0.875,0.5){$\delta_{2}$}
\put(1.875,0.5){$\delta_{3}$}
\put(2.875,0.5){$\delta_{4}$}
\put(3.875,0.5){$\delta_{5}$}
\put(1.5,2){$\delta_{6}$}
\put(1.875,3.5){$\delta_{0}$}
\put(5,1.5){$=$}
\put(6,1){\line(1,0){1}}
\multiput(7,1)(0.2,0){10}{\line(1,0){0.1}}
\put(9,1){\line(1,0){1}}
\multiput(8,1)(0,0.2){5}{\line(0,1){0.1}}
\put(8,2){\line(0,1){1}}
\multiput(6,1)(1,0){5}{\circle*{0.2}}
\put(8,2){\circle*{0.2}}
\put(8,3){\circle*{0.2}}
\put(5.875,0.5){$\alpha_{2}$}
\put(6.875,0.5){$\alpha_{1}$}
\put(7.875,0.5){${\bf w}$}
\put(8.875,0.5){$\beta_{1}$}
\put(9.875,0.5){$\beta_{2}$}
\put(7.5,2){$\gamma_{1}$}
\put(7.875,3.5){$\gamma_{2}$}
\end{picture}
\end{center}
\caption{
The extended Dynkin diagramm of $E_{6}$ is used to display the relations
between the simple roots of $E_{6}$ and those of the
subalgebra $A_{2}^{\oplus 3}$.}
\label{EsixBasisAtwo}
\end{figure}

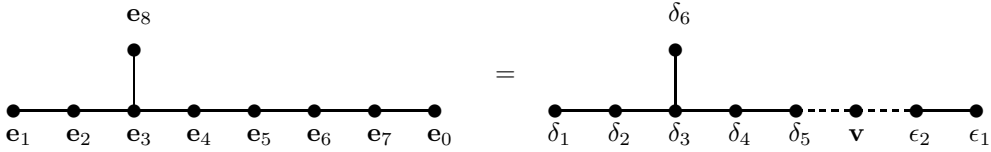
\begin{figure}
\unitlength0.8cm
\begin{picture}(16,3)
\put(0,1){\line(1,0){7}}
\put(2,1){\line(0,1){1}}
\multiput(0,1)(1,0){8}{\circle*{0.2}}
\put(2,2){\circle*{0.2}}
\put(-0.125,0.5){${\bf e}_{1}$}
\put(0.875,0.5){${\bf e}_{2}$}
\put(1.875,0.5){${\bf e}_{3}$}
\put(2.875,0.5){${\bf e}_{4}$}
\put(3.875,0.5){${\bf e}_{5}$}
\put(4.875,0.5){${\bf e}_{6}$}
\put(5.875,0.5){${\bf e}_{7}$}
\put(6.875,0.5){${\bf e}_{0}$}
\put(1.875,2.5){${\bf e}_{8}$}
\put(8,1.5){$=$}
\put(9,1){\line(1,0){4}}
\multiput(13,1)(0.2,0){10}{\line(1,0){0.1} }
\put(15,1){\line(1,0){1}}
\put(11,1){\line(0,1){1}}
\multiput(9,1)(1,0){8}{\circle*{0.2}}
\put(11,2){\circle*{0.2}}
\put(8.875,0.5){$\delta_{1}$}
\put(9.875,0.5){$\delta_{2}$}
\put(10.875,0.5){$\delta_{3}$}
\put(11.875,0.5){$\delta_{4}$}
\put(12.875,0.5){$\delta_{5}$}
\put(13.875,0.5){${\bf v}$}
\put(14.875,0.5){$\epsilon_{2}$}
\put(15.875,0.5){$\epsilon_{1}$}
\put(10.875,2.5){$\delta_{6}$}
\end{picture}
\caption{
The extended Dynkin diagramm of $E_{8}$ is used to display the
relations between the simple roots of $E_{8}$ and those of the
subalgebra $E_{6} \oplus A_{2}$.}
\label{EeightBasisEsixAtwo}
\end{figure}

Since the gauge twist is defined by its action on a
$A_{2}^{\oplus 3}$ sublattice of one $E_{8}$ lattice, we will
ignore the other $E_{8}$. We will use the following chain of
maximal regular semisimple subalgebras\footnote{The relevant
facts about subalgebras of simple Lie algebras can be found in
\cite{Cah}. For applications to symmetry breaking in toroidal
compactifications, see \cite{Moh}.}
(and the corresponding
root lattices):
\be     A_{2}^{\alpha} \oplus A_{2}^{\beta} \oplus
        A_{2}^{\gamma} \oplus A_{2}^{\epsilon}
        \subset
        E_{6}^{\delta} \oplus  A_{2}^{\epsilon}
        \subset
        E_{8}.
\label{embeddings}
\eq
Each subalgebra is labled with some letter, which is also used to
lable the simple roots, which spann the corresponding root lattice.
Thus $\alpha_{i}, \beta_{i}, \gamma_{i}, \epsilon_{i}$,
$i=1,2$ are simple roots of four (pairwise orthogonal)
$A_{2}$ subalgebras, $\delta_{j}$, $j=1,\ldots,6$ are simple roots
of $E_{6}^{\delta}$ and ${\bf e}_{k}$, $k=1,\ldots,8$ are simple
roots of the $E_{8}$. We will also need the lowest roots
$\delta_{0}$, ${\bf e}_{0}$ of $E_{6}^{\delta}$ and $E_{8}$.
The corresponding fundamental weights are denoted by
$\alpha_{i}^{*}, \ldots$. They are by definition dual
to the roots:
\be     \alpha_{i}^{*} \cdot \alpha_{j} = \delta_{ij},\ldots
\eq
All algebraic relations between the roots\footnote{To express
the fundamental weights in terms of the roots, one has to use
the inverse Cartan matrix.}
are summarized by
the extended root diagramms of $E_{6}^{\delta}$ and $E_{8}$.
See figures \ref{EsixBasisAtwo} and \ref{EeightBasisEsixAtwo}.

These diagramms make transparent, how the embeddings
(\ref{embeddings}) work: The vectors
\be     {\bf w} := \delta_{3} = -\alpha_{1}^{*}
        -\beta_{1}^{*} - \gamma_{1}^{*}
\eq
\be     {\bf v} := {\bf e}_{6} = -\delta_{5}^{*} - \epsilon_{2}^{*}
        = - \beta_{2}^{*} + \gamma_{2}^{*} - \epsilon_{2}^{*}
\eq
play a special role, because they ``glue together'' the semisimple
Lie algebras $A_{2}^{\oplus 3}$, $E_{6} \oplus A_{2}$
to bigger simple ones ($E_{6}$, $E_{8}$).

\subsection{Parametrization of the twist}

The gauge twist will be realized as a simultanuous Coxeter
twist in $A_{2}^{\alpha}$, $A_{2}^{\beta}$ and $A_{2}^{\gamma}$
whereas $A_{2}^{\epsilon}$ (and of cause the second $E_{8}$)
is invariant.
The action of the twist on the simple roots of $A_{2}^{\alpha}$
is given by
\be     \alpha_{1} \longrightarrow \alpha_{2}
        \longrightarrow - \alpha_{1} - \alpha_{2} \longrightarrow
        \alpha_{1} \rightarrow \cdots
\eq
The six simple roots $\pm \alpha_{1}$, $\pm\alpha_{2}$ und
$\pm (\alpha_{1} + \alpha_{2})$ form two orbits.
Equivalent formulas hold for $A_{2}^{\beta}$ and $A_{2}^{\gamma}$.
The fundamental weights $\alpha_{1}^{*}$, $\alpha_{2}^{*}$ are the
highest weights of the ${\bf 3}$, $\ov{\bf 3}$ representations
respectively. Their orbits under the twist give the other
weights of these representations:
\be     \alpha_{1}^{*}
        \longrightarrow \alpha_{2}^{*} - \alpha_{1}^{*}
        \longrightarrow - \alpha_{2}^{*}
        \longrightarrow \alpha_{1}^{*} \rightarrow \cdots
\eq
\be     \alpha_{2}^{*}
        \longrightarrow \alpha_{1}^{*} - \alpha_{2}^{*}
        \longrightarrow - \alpha_{1}^{*}
        \longrightarrow \alpha_{2}^{*} \rightarrow \cdots
\eq
As we have expressed the simple roots of the $E_{6}^{\delta}$
and $E_{8}$ in terms of the roots and weights of
the $A_{2}^{\oplus 4}$ subalgebra, their behaviour under the
twist is also clear.

\subsection{Parametrization of the Wilson lines}

According to our general results, there will be six continuous
Wilson lines $\wh{\bf A}_{i}$, $i=1,\ldots,6$, taking values in
the sixdimensional space spanned by the
$A_{2}^{\alpha} \oplus  A_{2}^{\beta} \oplus  A_{2}^{\gamma}$
lattice:
\be     \wh{V} = \la
        A_{2}^{\alpha} \oplus  A_{2}^{\beta} \oplus  A_{2}^{\gamma}
        \ra_{\fs{R}} \subset \la E_{8} \oplus E_{8} \ra_{\fs{R}}
\eq
We will postpone the analysis of discrete Wilson lines, setting
${\bf A}_{i} = \wh{\bf A}_{i}$, $\wt{\bf A}_{i} =0$.
The Wilson lines must therefore strictly transform like the
corresponding directions in target space and by the lifting
prescription like the six simple roots of the $A_{2}^{\oplus 3}$
lattice:
\be     {\bf A}_{2i-1} \longrightarrow {\bf A}_{2i}
        \longrightarrow
        -  {\bf A}_{2i-1}  -  {\bf A}_{2i} \longrightarrow
        {\bf A}_{2i-1} \longrightarrow
        \cdots
        \;\;\; i=1,2,3.
\eq
Our formulas precdict that there are nine metric and nine
isometric Wilson moduli. These are easily identified.
The geometric meaning of the metric Wilson moduli is the same
as that of the metric moduli: One can scale three of
six Wilson lines and can rotate the planes spanned by the
three pairs of Wilson lines relative to each other. Thus one
parametrization of the metric Wilson moduli is
given by $|{\bf A}_{1}|$, $|{\bf A}_{3}|$, $|{\bf A}_{5}|$,
$({\bf A}_{1} \cdot {\bf A}_{3})$,
$({\bf A}_{1} \cdot {\bf A}_{4})$,
$({\bf A}_{1} \cdot {\bf A}_{5})$,
$({\bf A}_{1} \cdot {\bf A}_{6})$,
$({\bf A}_{3} \cdot {\bf A}_{5})$,
$({\bf A}_{3} \cdot {\bf A}_{6})$.
The nine isometric Wilson moduli parametrize all rigid rotations
of the six Wilson lines in $\wh{V}$ that commute with the
twist.

In this paper we will only study the nine metric Wilson moduli
and keep the nine isometric Wilson moduli fixed. It will be
convenient to parametrize the Wilson lines by
weights of the $A_{2}^{\oplus 3}$ subalgebra in order to
control the breaking of the gauge group. The coordinates
on the nine dimensional subspace of the moduli space
are the nine real numbers $p,q,r$, $x,y,z$, $a,b,c$, defined by:
\be     {\bf A}_{1} = p \alpha_{1}^{*} + q \beta_{1}^{*}
                + r \gamma_{1}^{*},
\label{WL1}
\eq
\be     {\bf A}_{2} = p (\alpha_{2}^{*} - \alpha_{1}^{*})
                    + q (\beta_{2}^{*}  - \beta_{1}^{*})
                    + r (\gamma_{2}^{*}  - \gamma_{1}^{*}),
\eq
\be     {\bf A}_{3} = x \alpha_{1}^{*} + y \beta_{1}^{*}
                      + z \gamma_{1}^{*},
\eq
\be     {\bf A}_{4} = x (\alpha_{2}^{*} - \alpha_{1}^{*})
                    + y (\beta_{2}^{*}  - \beta_{1}^{*})
                    + z (\gamma_{2}^{*}  - \gamma_{1}^{*}),
\eq
\be     {\bf A}_{5} = a \alpha_{1}^{*} + b \beta_{1}^{*}
                      + c \gamma_{1}^{*},
\eq
\be     {\bf A}_{6} = a (\alpha_{2}^{*} - \alpha_{1}^{*})
                    + b (\beta_{2}^{*}  - \beta_{1}^{*})
                    + c (\gamma_{2}^{*}  - \gamma_{1}^{*}).
\label{WL6}
\eq

\subsection{A Strategy for the determination of the gauge group in
orbifold models with continuous Wilson lines}

The gauge group of an orbifold model can be calculated in two steps.
First one must know the gauge group of the underlying toroidal
compactification as a function of the moduli, then one has
to apply the twist in order to calculate the invariant subgroup.
Whereas much is known about symmetry breaking and
symmetry enhancement by Wilson lines in toroidal models \cite{Moh},
the explicit construction of the twist invariant subgroup is tedious and
difficult. We will therefore try to determine it indirectly by
combining our knowledge about the underlying torus compactification
with some general facts about Lie algebras and their automorphisms.
This will turn out to be sufficent in our example.

The toroidal model is completely specified by its momentum
lattice. We will denote by $\Gamma$ the momentum lattice with
no Wilson lines and generic (twist compatible) metric and
axionic moduli. Its gauge Lie algebra will be
$E_{8} \oplus E_{8} \oplus u(1)^{6}$, where the $u(1)^{6}$
comes from the six leftmoving internal oszillators $\alpha^{i}_{-1}$.
Since our target space twist has no invariant subspace, this
$u(1)^{6}$ will always be broken by the twist. Therefore it will be
ignored in the following. The gauge twist is known\footnote{We will
rederive this result in our formalim later.} to break
the $E_{8} \oplus E_{8}$ to a $E_{6} \oplus A_{2} \oplus E_{8}$.
If we switch on the nine twist compatible metric Wilson moduli,
the original momentum lattice $\Gamma$ is deformed into another
momentum lattice $\Gamma'$. It is well known how to control
what happens to the root vectors of the $E_{8} \oplus E_{8}$:
To each root ${\bf e}$ of $E_{8} \oplus E_{8}$, there corresponds
a vector $\wh{\bf e}$ in $\la \Gamma' \ra_{\fs{R}}$,
\be     \wh{\bf e} = ({\bf e}, {\bf 0}_{d}; {\bf 0}_{d}),
\eq
which is a lattice vector  of $\Gamma'$ if and only if it is
an integer linear combinations of the lattice basis of $\Gamma'$.
As $\Gamma'$ is selfdual, this is equivalent to $\wh{\bf e}$
having integer scalar products with all basis vectors
(\ref{basisvector_k}) - (\ref{basisvector_l}).
The unbroken
gauge Lie algebra of the deformed torus model is determined by
the subset of roots that remains in the lattice. As shown in
\cite{Moh} it is sufficent to control the simple roots.
For generic values, all $E_{6}^{\delta}$  roots will not be in
the deformed lattice. Thus the first $E_{8}$ is broken to
$u(1)^{6} \oplus A_{2}^{\epsilon}$. Since the $u(1)^{6}$ comes from six
internal oscillators, which transform non trivial under the gauge
twist, this will be further broken to $A_{2}^{\epsilon}$
in the corresponding orbifold model. (The second $E_{8}$ can only
be broken by discrete Wilson lines.)

To analyze the symmetry breakings induced by the nine metric Wilson
moduli systematically,
we will proceed in three steps. First we will study the
effect on the sublattice
$A_{2}^{\alpha} \oplus A_{2}^{\beta}  \oplus A_{2}^{\gamma}$,
then on $E_{6}^{\delta}$ and then on the complete first $E_{8}$.

\subsubsection{Step 1: Wilson lines on $A_{2}^{\oplus 3}$}

The vectors $\wh{\alpha}_{i}$
corresponding to the simple roots $\alpha_{i}$,
$i=1,2$ of $A_{2}^{\alpha}$  are elements of the
lattice $\Gamma'$ if and only if the parameters $p,x,a$ take
simultanously integer values. This is seen by working out the
scalar products between them and the lattice basis (\ref{basisvector_k})
- (\ref{basisvector_l}) using the
parametrization (\ref{WL1}) - (\ref{WL6})
of the Wilson lines. If these three conditions
are fulfilled, the gauge symmetry in the (corresponding sector of the)
toroidal model is $A_{2}$ instead of $u(1)^{2}$. The six states
corresponding to the six roots can be combined into eigenstates
of the twist. Two of these combinations, namely the orbits of
$|\alpha_{1} \ra$ and $|-\alpha_{1} \ra$ under the twist are
invariant and therefore they are not projected out by the twist.
Thus the twist breaks the $A_{2}$ to a $u(1)^{2}$.
The dimension of the subspace where this happens is six, because we
have to set three moduli to integer values. The corresponding
critical Wilson lines will be called type--$\alpha$ Wilson lines.

Generalizing to the complete $A_{2}^{\oplus 3}$, there are also
type--$\beta$ and type--$\gamma$ Wilson lines:
\be
\begin{array}{lllll}
        \mbox{type--} \alpha & :\Leftrightarrow  & \wh{\alpha}_{i}
        \in \Gamma'& \Leftrightarrow
        & p,x,a \in \Zzahl    \\
        \mbox{type--} \beta & :\Leftrightarrow  & \wh{\beta}_{i}
        \in \Gamma'& \Leftrightarrow
        & q,y,b \in \Zzahl   \\
        \mbox{type--} \gamma & :\Leftrightarrow   & \wh{\gamma}_{i}
        \in \Gamma' & \Leftrightarrow
        & r,z,c \in \Zzahl  \\
\end{array}
\eq
Combining the different conditions, we define type--($\alpha, \beta$)
Wilson lines and so on. Summarizing all possible combinations
we arrive at table \ref{A23}. Geometrically the three critical
three--dimensional subspaces intersect each other in discrete
maximal critical points, and they are themselves the intersection
loci of the three critical six--dimensional subspaces.

\begin{table}
\[
\begin{array}{|c|c|c|c|} \hline
\mbox{type} & \mbox{Lie algebra, torus} & \mbox{Lie algebra, orbifold}
        &\mbox{moduli} \\   \hline \hline
\alpha, \beta, \gamma & A_{2} \oplus A_{2} \oplus A_{2} &
u(1)^{6} & 0 \\ \hline
\alpha, \beta & A_{2} \oplus A_{2} \oplus u(1)^{2} &
u(1)^{4} & 3 \\ \hline
\alpha, \gamma & A_{2}  \oplus u(1)^{2} \oplus A_{2} &
u(1)^{4} & 3 \\ \hline
\beta, \gamma & u(1)^{2} \oplus A_{2}  \oplus  A_{2} &
u(1)^{4} & 3 \\ \hline
\alpha & A_{2} \oplus u(1)^{4} & u(1)^{2} & 6 \\ \hline
\beta & u(1)^{2} \oplus A_{2} \oplus u(1)^{2} & u(1)^{2} & 6 \\ \hline
\gamma & u(1)^{4} \oplus A_{2} & u(1)^{2} & 6 \\ \hline
- & u(1)^{6} & - & 9 \\ \hline
\end{array}
\]
\caption{Summary of the effect of metric Wilson moduli on the
$A_{2}^{\oplus 3}$ subalgebra. We display the type of the
Wilson line, as defined in the text, the gauge symmetries of
the toroidal and of the orbifold model, and the number of
those moduli, which can still be varied.}
\label{A23}
\end{table}

\subsubsection{Step 2: Wilson lines on $E_{6}^{\delta}$}

In order to pass from the $A_{2}^{\oplus 3}$ subalgebra
to the $E_{6}^{\delta}$, we have to study the effects of
the Wilson lines and the twist on the simple root ${\bf w}$.
Wilson lines, which are compatible with the corresponding
vector $\wh{\bf w}$ of the momentum lattice, will be called
type--${\bf w}$ Wilson lines:
\be     \mbox{type--${\bf w}$} :\Leftrightarrow
        \wh{\bf w} \in \Gamma' \Leftrightarrow
        p+q+r \in 3 \Zzahl \mbox{  and  }
        x+y+z \in 3 \Zzahl \mbox{  and  }
        a+b+c \in 3 \Zzahl.
\eq
If the Wilson lines are not type--${\bf w}$, then the
symmetry of the toroidal model is at most
$A_{2}^{\oplus 3}$ and the analysis of the last subsection
applies. Let us then look at Wilson lines, which are
type--${\bf w}$, but not type--$\alpha$, -$\beta$ or -$\gamma$.
In this case the vectors $\pm \wh{\bf w}$ are in $\Gamma'$.
But the Wilson lines are compatible with the gauge twist, and
all vectors belonging to the orbits of the two vectors must also
be in $\Gamma'$. These are, expressed as weights of the
$A_{2}^{\oplus 3}$ subalgebra:
\be     \pm(-\wh{\alpha}_{1}^{*} - \wh{\beta}_{1}^{*}
        - \wh{\gamma}_{1}^{*}), \;\;\;
        \pm(\wh{\alpha}_{1}^{*} - \wh{\alpha}_{2}^{*}
           + \wh{\beta}_{1}^{*} - \wh{\beta}_{2}^{*}
           + \wh{\gamma}_{1}^{*}- \wh{\gamma}_{2}^{*}  ), \;\;\;
        \pm(\wh{\alpha}_{2}^{*} + \wh{\beta}_{2}^{*}
        + \wh{\gamma}_{2}^{*}).
\eq
Since they form the root system of a $A_{2}$, the symmetry of
the torus model is enlarged from $u(1)^{6}$ to $A_{2} \oplus u(1)^{4}$.
In the corresponding orbifold model the residual symmetry will be
$u(1)^{2}$, because there are two twist invariant orbits from
the six roots.

If the Wilson lines are type--$({\bf w}, \alpha)$ the symmetry
is enlarged further. Beside the $2 \cdot 6$ roots from the
twist orbits of $\pm \alpha_{1}$ and $\pm {\bf w}$, we can
construct another 12 roots. This is most easily seen by observing
that all vectors in the orbits of $\pm {\bf w}$ contain a weight
of the $\ov{\bf 3}$ or ${\bf 3}$ representations of $A_{2}^{\alpha}$.
By adding a suitable root of $A_{2}^{\alpha}$ this weight can
be replaced by any other weight of the same representation
without changing the $A_{2}^{\beta} \oplus A_{2}^{\gamma}$ part
and the norm of the vector. Altogether there are 24 roots.
The resulting Lie algebra must have rang 6 and be a regular
subalgebra of $E_{6}^{\delta}$. Using Dynkins algorithm
for regular subalgebras, one can show that there is only one
such subalgebra, $D_{4} \oplus u(1)^{2}$. This is the
enlarged symmetry of the torus model. The orbifold twist
will break the $u(1)^{2}$ completely. The $D_{4}$ roots can
be arranged into 8 invariant orbits. This counting shows that
the Lie algebra of the twisted model is $A_{2}$ or
$A_{1} \oplus A_{1} \oplus u(1)^{2}$. If the twist is an
inner automorphism of $D_{4}$, it cannot reduce the rank
and the $A_{2}$ is excluded. However there is no reason for the
twist to be an inner automorphism: A twist of the
root lattice of a Lie algebra does only define an inner automorphism
if it is a Weyl twist. The outer automorphisms (up to inner ones)
are in one to one correspondence with symmetries of the Dynkin
diagram \cite{Cah}. The Lie algebra $D_{4}$ is the only simple
Lie algebra with an outer automorphim of order three. As all
other simple Lie algebra have outer automorphisms of at most order two,
and we are studying a twist of order three, we do not know
whether the automorphism is inner or outer if and only if
the symmetry of the toroidal model is $D_{4}$.

Fortunately we can determine the rank of the twist invariant
subalgebra by writing down all eight invariant combinations
of Kac--Moody currents and calculating their operator product expansions.
Then we see that the currents corresponding to the orbits
of $\pm \alpha_{1}$ fulfill a $u(1)^{2}$ current algebra, and that
their OPEs with the other six allways contain a first order pole.
This shows that the rank of the orbifold Lie algebra is two.
Thus the algebra is $A_{2}$ and the automorphism must be an outer one.

The cases of type--$({\bf w}, \beta)$ and type--$({\bf w}, \gamma)$
Wilson lines are completely analog. The next step is to note
that there are no type--$({\bf w}, \alpha, \beta)$,
type--$({\bf w}, \alpha, \gamma)$ or type--$({\bf w}, \beta, \gamma)$
Wilson lines, because the corresponding conditions cannot be
fulfilled simultanously. There are however
type--$({\bf w}, \alpha, \beta, \gamma)$ Wilson lines. In this case all
simple roots of $E_{6}^{\delta}$ are present. (Looking into the
details, we find, that by adding $A_{2}^{\oplus 3}$ roots to the
vectors in the orbits of $\pm{\bf w}$, all $2 \cdot 27$ weights
of the representations $(\ov{\bf 3},\ov{\bf 3}, \ov{\bf 3})$
$({\bf 3},{\bf 3}, {\bf 3})$ can be constructed, thus extending
the $A_{2}^{\oplus 3}$ root lattice to the $E_{6}$ root lattice.)
To calculate the orbifold Lie algebra, it is sufficent to count
the invariant orbits: Since there are 24 invariant states, and the
automorphism must be inner, we must look for a regular rank six
subalgebra of $E_{6}$ with dimension 24. Using Dynkins formalism
we see that the unique solution is $A_{2}^{\oplus 3}$.

Summarizing, the possible breakings of $E_{6}$ by twist and
continuous Wilson lines are collected in table \ref{E6}.
The number of residual moduli is again calculated by counting
the conditions, that the Wilson moduli must fulfill.
\begin{table}
\[
\begin{array}{|c|c|c|c|}   \hline
\mbox{Type} & \mbox{Lie algebra, torus} & \mbox{Lie algebra, orbifold}
        & \mbox{Moduli} \\  \hline \hline
{\bf w}, \alpha, \beta, \gamma &
E_{6} & A_{2} \oplus A_{2} \oplus A_{2} & 0 \\ \hline
{\bf w}, \alpha & D_{4} \oplus u(1)^{2} & A_{2} & 3 \\  \hline
{\bf w}, \beta & D_{4} \oplus u(1)^{2} & A_{2} & 3 \\  \hline
{\bf w}, \gamma & D_{4} \oplus u(1)^{2} & A_{2} & 3 \\  \hline
{\bf w} & A_{2} \oplus u(1)^{4} & u(1)^{2} & 6 \\   \hline
\end{array}
\]
\caption{Summary of the effect of type--${\bf w}$ Wilson lines
on the $E_{6}$ subalgebra. For Wilson lines, which are not
type--${\bf w}$, see table 1.}
\label{E6}
\end{table}

\subsubsection{Step 3: Wilson lines on $E_{8}$}

The last step is to extend our analysis to the full $E_{8}$ lattice.
As the sublattice $A_{2}^{\epsilon}$ is both twist invariant
and orthogonal to the Wilson lines, the corresponding $A_{2}$
algebra will be unbroken in the toroidal and in the orbifold model.
The only thing to do is to look at the lattice vector $\wh{\bf v}$
corresponding to the simple root ${\bf v}$ of $E_{8}$ which
extends $E_{6}^{\delta} \oplus A_{2}^{\epsilon}$ to $E_{8}$.
This leeds to a new type of critical Wilson lines:
\be     \mbox{type--{\bf v}} :\Leftrightarrow
        \wh{\bf v} \in \Gamma' \Leftrightarrow
        (q-r) \in 3 \Zzahl \mbox{  and  }
        (y-z) \in 3 \Zzahl \mbox{  and  }
        (b-c) \in 3 \Zzahl.
\eq
Using the same methods as in the last subsection, one arrives
at table \ref{E8}.
Instead of going through all the cases, let us comment on the
case of type--${\bf v}$ Wilson lines. Here it is easily shown, that
the unbroken Lie algebra of the torus model is $u(1)^{4}  \oplus D_{4}$.
By counting the invariant states we know that the Lie algebra of
the corresponding orbifold has dimension 14. We also know that
it has $A_{2}^{\epsilon}$ as a subalgebra and that the rank
is between two and four.
Using Dynkins algorithm it is easy to show, that there is no
regular subalgebra of $D_{4}$ with these properties. It is
however well know that $D_{4}$ has a non--regular
(and also non simply laced) subalgebra $G_{2}$, which is the
invariant subalgebra under a third order outer
automorphism\footnote{This is used to construct the
twisted affine Lie algebra $D_{4}^{(3)}$ \cite{Kac}.}.
Therefore, this must be the unbroken Lie algebra in this case.
All other cases in which similar complications arrise can be
reduced to this case. For a summary see table \ref{E8}.

\subsection{The Effect of discrete Wilson lines}

Our last point is to study the effect of discrete Wilson lines
on the gauge group. Again we will restrict our attention to the
first $E_{8}$. According to the general results there are two
constraints for discrete Wilson lines $\wt{\bf A}_{i}$:
The first is, that multiplying $\wt{\bf A}_{i}$ with the
order of the twist must give an invariant lattice vector of
the $E_{8}$ lattice. The sublattice of (pointwise)
invariant lattice vectors is precisely the root lattice
$A_{2}^{\epsilon}$:
\be     3 \wt{\bf A}_{i} \in I_{E_{8}} = A_{2}^{\epsilon}
\eq
The other constraint is, that the  $\wt{\bf A}_{i}$ must transform,
up to a lattice vector of $I_{E_{8}}$, as the corresponding
directions transform under the target space twist:
\be   \theta' \wt{\bf A}_{i}  -  \theta_{i}^{\;j} \wt{\bf A}_{j} =
      (\delta_{i}^{\;j}  -  \theta_{i}^{\;j}) \wt{\bf A}_{j}
        \in  A_{2}^{\epsilon}.
\eq
Using the explicit form of the twist we get
\be   \left\{  \begin{array}{l}  \wt{\bf A}_{2k-1} - \wt{\bf A}_{2k}
                        \in A_{2}^{\epsilon},  \\
                \wt{\bf A}_{2k-1} +  2 \wt{\bf A}_{2k}  \in
                         A_{2}^{\epsilon},  \\
        \end{array}  \right.
\eq
with $k=1,2,3$. Rearranging this to
\be    \left\{  \begin{array}{ll}
        3 \wt{\bf A}_{i}  \in A_{2}^{\epsilon},& i=1,\ldots,6, \\
        \wt{\bf A}_{2k-1} - \wt{\bf A}_{2k}
        \in A_{2}^{\epsilon}, & k=1,2,3,\\
        \end{array}  \right.
\eq
we learn that those discrete Wilson lines,
which are connected by the target
space twist,
must be in the same coset of $\frac{1}{3} A_{2}^{\epsilon}$
with respect to $A_{2}^{\epsilon}$. There are nine such cosets.
Their canonical representatives are
\be     \frac{m}{3} \epsilon_{1} + \frac{n}{3} \epsilon_{2},
        \;\;\; m,n = 0,1,2.
\eq
To determine the effect on the gauge group, we note that
three of these are weights of the Lie algebra $A_{2}^{\epsilon}$,
namely
\be    \frac{2}{3} \alpha_{1} + \frac{1}{3} \alpha_{2} = \alpha_{1}^{*},
        \;\;\;
        \frac{1}{3} \alpha_{1} + \frac{2}{3} \alpha_{2} = \alpha_{2}^{*}
        \mbox{  und  } {\bf 0}.
\eq
whereas the other six are not. Concerning the gauge group there
are only two possibilities: The first is that all six
Wilson lines $\wt{\bf A}_{i}$ are weights of $A_{2}^{\epsilon}$.
Then the $A_{2}^{\epsilon}$ roots are still present in the
deformed lattice $\Gamma'$. Thus the $A_{2}^{\epsilon}$ will
not be broken in the torus model, and since the twist
operates trivially on $A_{2}^{\epsilon}$, it is also unbroken
in the twisted model. The gauge groups are the same as in the
case without discrete Wilson lines and table \ref{E8} remains
valid. These Wilson lines will be called type--$\epsilon$.

But if at least one pair of Wilson lines takes values in a
different coset, then the $A_{2}^{\epsilon}$ is broken to a
$u(1)^{2}$ in the torus model. Again, this is not further broken
by the twist. In order to modify table \ref{E8} we have to look
at the vector ${\bf v}$, because it is a non--trivial
weight of $A_{2}^{\epsilon}$.
To derive a condition for $\wh{\bf v}$ to be in $\Gamma'$,
let us parametrize the discrete Wilson lines by their
canonical coset representatives:
\be     \wt{\bf A}_{1} \simeq \wt{\bf A}_{2} \simeq \frac{m_{1}}{3}
                \epsilon_{1} + \frac{n_{1}}{3} \epsilon_{2}, \;\;\;
        \wt{\bf A}_{3} \simeq \wt{\bf A}_{4} \simeq \frac{m_{2}}{3}
                \epsilon_{1} + \frac{n_{2}}{3} \epsilon_{2}, \;\;\;
        \wt{\bf A}_{5} \simeq \wt{\bf A}_{6} \simeq \frac{m_{3}}{3}
                \epsilon_{1} + \frac{n_{3}}{3} \epsilon_{2}.
\eq
The modified definition of a type--${\bf v}$ Wilson line for
nontrivial discrete Wilson lines is:
\be     \mbox{type--{\bf v}} :\Leftrightarrow
        \wh{\bf v} \in \Gamma'
\eq
\[   \Leftrightarrow (q-r-n_{1}) \in 3 \Zzahl \mbox{  and  }
     (y-z-n_{2})     \in 3 \Zzahl \mbox{  and  }
     (b-c-n_{3})     \in 3 \Zzahl.
\]
We can now study the effect of the nine metric Wilson moduli
with the same methods as before. The results are shown
in table \ref{E8var}.

\begin{table}
\[
\begin{array}{|c|c|c|c|c|}   \hline
\mbox{Type} & \mbox{Lie algebra, torus}
& \mbox{Lie algebra, orbifold} &
        \mbox{gauge group} & \mbox{Moduli} \\  \hline \hline
{\bf v}, {\bf w},\alpha,\beta,\gamma &
E_{8} &  E_{6} \oplus A_{2} & E(6) \otimes SU(3) & 0 \\ \hline
{\bf v}, \alpha,\beta,\gamma & A_{2} \oplus E_{6} &
D_{4} \oplus u(1)^{4} & SO(8) \otimes U(1)^{2} & 0 \\ \hline
{\bf w}, \alpha,\beta, \gamma & E_{6} \oplus A_{2} &
A_{2} \oplus A_{2} \oplus A_{2} \oplus A_{2} &
(SU(3))^{4} & 0 \\  \hline
\alpha,\beta,\gamma & A_{2} \oplus A_{2} \oplus A_{2} \oplus A_{2} &
u(1)^{6} \oplus A_{2} & SU(3) \otimes U(1)^{6} & 0 \\ \hline
{\bf v}, {\bf w},\alpha & D_{4} \oplus D_{4} \oplus u(1)^{2} &
A_{2} \oplus G_{2} &G(2) \otimes SU(3) & 0 \\ \hline
{\bf v}, {\bf w} & A_{2} \oplus D_{4} \oplus u(1)^{2} &
G_{2} \oplus u(1)^{2} & G(2) \otimes U(1)^{2} & 3 \\ \hline
{\bf v}, \alpha & A_{2} \oplus D_{4} \oplus u(1)^{2} &
G_{2} \oplus u(1)^{2} & G(2) \otimes U(1)^{2} & 3 \\ \hline
{\bf v}, \beta, \gamma & u(1)^{2} \oplus E_{6} &
D_{4} \oplus u(1)^{2} & SO(8) \otimes U(1)^{2} & 3 \\ \hline
\alpha, \beta & A_{2} \oplus A_{2} \oplus u(1)^{2} \oplus A_{2} &
u(1)^{4} \oplus A_{2} & SU(3) \otimes U(1)^{4} & 3 \\ \hline
\alpha, \gamma & A_{2} \oplus A_{2} \oplus u(1)^{2} \oplus A_{2} &
u(1)^{4} \oplus A_{2} & SU(3) \otimes U(1)^{4} & 3 \\ \hline
\beta, \gamma & A_{2} \oplus A_{2} \oplus u(1)^{2} \oplus A_{2} &
u(1)^{4} \oplus A_{2} & SU(3) \otimes U(1)^{4} & 3 \\ \hline
{\bf w}, \alpha & D_{4} \oplus u(1)^{2} \oplus A_{2} &
A_{2} \oplus A_{2} & SU(3) \otimes SU(3) &3 \\ \hline
{\bf w}, \beta & D_{4} \oplus u(1)^{2} \oplus A_{2} &
A_{2} \oplus A_{2} & SU(3) \otimes SU(3) &3 \\ \hline
{\bf w}, \gamma & D_{4} \oplus u(1)^{2} \oplus A_{2} &
A_{2} \oplus A_{2} & SU(3) \otimes SU(3) &3 \\ \hline
{\bf w} & A_{2} \oplus u(1)^{4} \oplus A_{2} &
u(1)^{2} \oplus A_{2} & SU(3) \otimes U(1)^{2} & 6\\ \hline
\alpha & A_{2} \oplus u(1)^{4} \oplus A_{2} &
u(1)^{2} \oplus A_{2} & SU(3) \otimes U(1)^{2} & 6\\ \hline
\beta & A_{2} \oplus u(1)^{4} \oplus A_{2} &
u(1)^{2} \oplus A_{2} & SU(3) \otimes U(1)^{2} & 6\\ \hline
\gamma & A_{2} \oplus u(1)^{4} \oplus A_{2} &
u(1)^{2} \oplus A_{2} & SU(3) \otimes U(1)^{2} & 6\\ \hline
{\bf v} & D_{4} \oplus u(1)^{4} & G_{2} & G(2)& 6 \\ \hline
- & u(1)^{6} \oplus A_{2} & A_{2} & SU(3) & 9 \\ \hline
\end{array}
\]
\caption{
In this table we summarize all possibilities of
breaking the standard $\Zzahl_{3}$ orbifold by
nine metric Wilson moduli, but without discrete Wilson
lines. We display the type of critical Wilson lines as
explained in the text, the Lie algebras realized in the
untwisted (toroidal) and twisted (orbifold) model,
the unbroken gauge group of the orbifold model, and the number
of moduli, that can still be varied continuously.
}

\label{E8}
\end{table}

\begin{table}
\[
\begin{array}{|c|c|c|c|c|}   \hline
\mbox{Type} & \mbox{Lie  Algebra, torus}
& \mbox{Lie Algebra, orbifold} &
        \mbox{gauge group} & \mbox{Moduli} \\  \hline \hline
{\bf v}, {\bf w},\alpha,\beta,\gamma &
E_{7}\oplus u(1) &  A_{5} \oplus A_{2} \oplus u(1) &
SU(6) \otimes SU(3) \otimes U(1) & 0 \\ \hline
{\bf v}, \alpha,\beta,\gamma & A_{2} \oplus A_{5} \oplus u(1) &
u(1)^{6} \oplus A_{2} & SU(3) \otimes U(1)^{6} & 0 \\ \hline
{\bf w}, \alpha,\beta, \gamma & E_{6} \oplus u(1)^{2} &
A_{2} \oplus A_{2} \oplus A_{2} \oplus u(1)^{2} &
(SU(3))^{3}\otimes U(1)^{2} & 0 \\  \hline
\alpha,\beta,\gamma & A_{2} \oplus A_{2} \oplus A_{2} \oplus u(1)^{2} &
u(1)^{8}  & U(1)^{8} & 0 \\ \hline
{\bf v}, {\bf w},\alpha & D_{4} \oplus A_{2} \oplus u(1)^{4} &
A_{2} \oplus u(1)^{2} &SU(3) \otimes U(1)^{2} & 0 \\ \hline
{\bf v}, {\bf w} & A_{2} \oplus A_{2} \oplus u(1)^{4} &
u(1)^{4} & U(1)^{4} & 3 \\ \hline
{\bf v}, \alpha & A_{2} \oplus A_{2} \oplus u(1)^{4} &
u(1)^{4} & U(1)^{4} & 3 \\ \hline
{\bf v}, \beta, \gamma & u(1)^{3} \oplus A_{5} &
A_{2} \oplus u(1)^{2} & SU(3) \otimes U(1)^{2} & 3 \\ \hline
\alpha, \beta & A_{2} \oplus A_{2} \oplus u(1)^{4}  &
u(1)^{6}  & U(1)^{6} & 3 \\ \hline
\alpha, \gamma & A_{2} \oplus A_{2} \oplus u(1)^{4}  &
u(1)^{6} & U(1)^{6} & 3 \\ \hline
\beta, \gamma & A_{2} \oplus A_{2} \oplus u(1)^{4}  &
u(1)^{6} &  U(1)^{6} & 3 \\ \hline
{\bf w}, \alpha & D_{4} \oplus u(1)^{4}  &
A_{2} \oplus u(1)^{2} & SU(3) \otimes U(1)^{2} &3 \\ \hline
{\bf w}, \beta & D_{4} \oplus u(1)^{4}  &
A_{2} \oplus u(1)^{2} & SU(3) \otimes U(1)^{2} &3 \\ \hline
{\bf w}, \gamma & D_{4} \oplus u(1)^{2} \oplus u(1)^{2} &
A_{2} \oplus u(1)^{2} & SU(3) \otimes U(1)^{2} &3 \\ \hline
{\bf w} & A_{2} \oplus u(1)^{6} &
u(1)^{4}  &U(1)^{4} & 6\\ \hline
\alpha & A_{2} \oplus u(1)^{6}  &
u(1)^{4}  &  U(1)^{4} & 6\\ \hline
\beta & A_{2} \oplus u(1)^{6}  &
u(1)^{4} &  U(1)^{4} & 6\\ \hline
\gamma & A_{2} \oplus u(1)^{6} &
u(1)^{4}  & U(1)^{4} & 6\\ \hline
{\bf v} & A_{2} \oplus u(1)^{6} & u(1)^{4} & U(1)^{4} & 6 \\ \hline
- & u(1)^{8} & u(1)^{2} & U(1)^{2} & 9 \\ \hline
\end{array}
\]
\caption{
This table contains all possibilities of symmetry breaking
of the standard $\Zzahl_{3}$ orbifold in the case when
the subalgebra $A_{2}^{\epsilon}$ is broken by discrete Wilson
lines. The second $E_{8}$ is ignored.}
\label{E8var}
\end{table}

\section{Conclusions}

Summarizing, we have identified the untwisted moduli of
heterotic $\Zzahl_{N}$ orbifold compactifications and
developed a method to determine the gauge group, which was
sucessfully applied to the standard $\Zzahl_{3}$ orbifold.
All constraints on the background fields were derived
systematically from the basic constraint
$\Theta \in \mbox{Aut}(\Gamma_{16+d;d})$.
By separating the continuous and discrete parts of the background
fields, we have shown that the appearence of
discrete background fields is due to the fact
that the twist of the momentum lattice is not
uniquely fixed by the target space twist. The choice of the
gauge twist turned out to be highly restricted and the
lifting and embedding construction used here seems to be the
only natural solution.
The geometry of critical subspaces found in the $\Zzahl_{3}$ example
is similar to that found
in the case of toroidal compactifications in \cite{Moh}.
It should be emphasized, that the characterization
of critical subspaces is independent of
the determination of the gauge group. Thus even if it is
an accident that we could get the gauge groups of the orbifold
models without
explicit construction, the geometry of critical subspaces
should alway be accessible. And the result that the
critical geometries of the two orbifold moduli spaces connected
by choosing different discrete Wilson lines are in one to
one correspondence with one another and with the underlying
toroidal moduli spaces must also hold more generally.
The dimensions and critical subspaces
of different orbifold moduli spaces with the same
target space twist but inequivalent discrete background fields
should always be isomorphic.

There are also some implications concerning the modular groups
for orbifolds with Wilson moduli. Since points of maximaly extended
symmetry are usually fixpoints, we expect that critical subspaces
of non maximal symmetry also transform somewhat singular, for
example they could be invariant (not pointwise) or transform into
one another. The construction of modular groups will be one
of the next steps.

Note also that there are two further types of moduli,
which should be included in the future. For orbifolds without
Wilson lines it is known that there are moduli arrising from the
twisted sectors (twisted moduli) which parametrize the blowing up
of the orbifold singularities. If these can coexist with
the Wilson moduli, this could be used to explore more general
$(0,2)$ compactifications.
It is also possible that at points of extended symmetry
in the untwisted moduli space additional moduli arrise
which correspond to extra (non--redundant) marginal operators.
Since this is interelated with extended gauge symmetry and
special behaviour under modular transformations, thess moduli
should have a nice group theoretical characterisation, if they
exist.

There is of course a lot of other things that can be studied now,
for example the complete massless spectrum, the twisted sectors
(where the degeneracy of the ground states is lifted by the
Wilson lines \cite{INQ0}), the construction of
vertex operators, the calculation of correlators, and the
derivation of effective actions.

\end{document}